\documentclass[fleqn,usenatbib]{mnras}

\usepackage{newtxtext,newtxmath}

\usepackage[T1]{fontenc}
\usepackage{ae,aecompl}

\usepackage{graphicx}	
\usepackage{subfig}
\usepackage{amsmath}	
\usepackage{array}      
\usepackage{pdflscape}
\usepackage{rotating}
\usepackage{longtable}
\usepackage{verbatim}
\usepackage{pifont}
\usepackage{booktabs}

\usepackage[dvipsnames]{xcolor}

\newcommand{\gx}{GX~339--4}
\newcommand{\gr}{$\gamma$-ray}
\newcommand{\g}{$\gamma$}

\title[\gx\ in TeV \gr s]{The prototype X-ray binary \gx: using TeV \gr s to assess LMXBs as Galactic cosmic ray accelerators}

\author[D. Kantzas et al.]{
D. Kantzas$^{1,2}$\thanks{E-mail: d.kantzas@uva.nl}, 
S. Markoff$^{1,2}$, 
M. Lucchini$^{3,1}$,
C. Ceccobello$^4$,
V. Grinberg$^{5,6}$,
\newauthor
~R. M. T. Connors$^{7}$ \&
P. Uttley$^{1}$
\\
$^{1}$Anton Pannekoek Institute for Astronomy (API), University of Amsterdam, Science Park 904, 1098 XH Amsterdam, the Netherlands\\
$^{2}$GRavitational AstroParticle Physics Amsterdam (GRAPPA), University of Amsterdam, Science Park 904, 1098 XH Amsterdam, the Netherlands\\
$^3$MIT Kavli Institute for Astrophysics and Space Research, Massachusetts Institute of Technology, Cambridge, MA 02139, USA\\
$^4$Department of Space, Earth and Environment, Chalmers University of Technology, Onsala Space Observatory, 439 92 Onsala, Sweden\\
$^{5}$Institute for Astronomy und Astrophysics, University of T{\"u}bingen, Sand 1, 72076 T\"{u}bingen, Germany\\
$^6$European Space Agency (ESA), European Space Research and Technology Centre
(ESTEC), Keplerlaan 1, 2201 AZ Noordwijk, the Netherlands\\
$^{7}$Cahill Center for Astronomy and Astrophysics, California Institute of Technology, 1200 California Boulevard, Pasadena, CA 91125, USA
}

\date{Accepted XXX. Received YYY; in original form ZZZ}

\pubyear{2021}

\begin{document}
\label{firstpage}
\pagerange{\pageref{firstpage}--\pageref{lastpage}}
\maketitle

\begin{abstract}
Since the discovery of cosmic rays (CRs) over a century ago, their origin remains an open question. Galactic CRs with energy up to the knee ($10^{15}$\,eV) are considered to originate from supernova remnants, but this scenario has recently been questioned due to lack of TeV \gr\ counterparts in many cases. Extragalactic CRs on the other hand, are thought to be associated with accelerated particles in the relativistic jets launched by supermassive accreting black holes at the center of galaxies. Scaled down versions of such jets have been detected in X-ray binaries hosting a stellar black hole (BHXBs). In this work, we investigate the possibility that the smaller-scale jets in transient outbursts of low-mass BHXBs could be sources of Galactic CRs. To better test this scenario, we model the entire electromagnetic spectrum of such sources focusing on the potential TeV regime, using the 'canonical' low-mass BHXB \gx as a benchmark. Taking into account both the leptonic radiative processes and the \gr s produced via neutral pion decay from inelastic hadronic interactions, we predict the GeV and TeV \gr\ spectrum of \gx\ using lower-frequency emission as constraints. Based on this test-case of \gx\ we investigate whether other, nearby low-mass BHXBs could be detected by the next-generation very-high-energy \gr\ facility the Cherenkov Telescope Array,  which would establish them as additional and numerous potential sources of CRs in the Galaxy. 
\end{abstract}




\section{Introduction}\label{sec:intro}
Accreting supermassive black holes located at the centres of galaxies are the most powerful engines in the Universe, and some of the most interesting laboratories to investigate the physics of extreme gravity. Of particular importance are those active galactic nuclei (AGN) that exhibit relativistic and collimated jets. The underlying physics that unites the accretion of black holes with the large scale jets is still an unanswered problem. These relativistic jets are considered powerful enough to accelerate particles to very high energy, making them likely a source of extragalactic cosmic rays (CRs) that reach energies of at least $10^{19}\,$eV \citep{hillas1984origin,Abbasi_2018,Perrone_2020}. 

CRs are elementary particles and/or atoms of extraterrestrial origin. The resulting CR spectrum covers ten orders of magnitudes in particle energy and shows two very well known characteristic spectral features where the slope changes. The first one is the `knee' that is located around $10^{15}\,$eV, and the second feature is the `ankle' that is located around $10^{17}\,$eV. Current models assume that CRs up to the knee are produced within the Milky Way, while CRs from above the ankle are of extragalactic origin \citep{hillas1984origin,Drury2012origin,Blasi2013origin}. Supernova remnants have long been considered the dominant source of Galactic CRs based on their size and measured magnetic fields \citep{hillas1984origin,Volk2003variation,vink2012supernova,Ackermann2013detection}, but due to the lack of TeV \gr\ counterparts the debate is still open \citep{aharonian2018massive}. Given the ability of AGN jets to accelerate cosmic rays, another promising alternative source could be the Galactic jets launched in X-ray binaries comprised of a stellar accreting black hole and a companion star (BHXBs; \citealt{mirabel1994superluminal,fender2001powerful,mcclintock2006compact}). Such Galactic jets share the physical properties of AGN jets but on much smaller scales \citep[][]{heinz2002CRmicroquasars,romero2003hadronic,romero2005misaligned,fender2005CRXRBs,Romero2008proton,Vila2010gx339,Vila2012,pepe2015lepto,cooper2020xrbcrs,kantzas2020cyg}.

The presence of jets in low mass BHXBs is transient, and is tightly connected to the properties of the accretion flow. In the so-called hard state, BHXBs display a flat or inverted radio-to-IR spectrum associated with jet synchrotron emission analogous to AGN jets \citep{blandford1979relativistic,hjellming1988radio,falcke1995jet,markoff2001jet,fender2006transient,Corbel2003radio,Corbel2013universal}. BHXBs transit from quiescent to hard and soft states within `human-like' timescales, hence we can observe the jet launching and jet quenching in real-time \citep[see e.g.,][]{Russell_2020}. The dynamical timescales are roughly proportional to the mass of the black hole, so it would take typically millions of times longer to detect similar state transitions in AGN.

Accelerated particles in AGN jets are the source of the non-thermal radiation detected over the entire electromagnetic spectrum, from radio to TeV \gr s \citep[see e.g.][]{tavecchio1998constraints,Celotti2001,aharonian2004very,Georganopoulos_2006,marscher2008inner,Ghisellini2009}. However, the exact radiative mechanism has been under debate for a long time because it is tightly connected to the jet composition and the exact particle acceleration mechanism, which remain debated. Two scenarios are generally considered depending on the jet launching mechanism. First, a purely leptonic jet powered by the black hole spin \citep{blandford1977extraction} may accelerate electrons/positrons that are responsible for the entire multi-wavelength spectrum \citep{Maraschi1992,Dermer1993model,Levinson1995,Blandford1995pair,Marcowith1995beam,bottcher1997pair,Georganopoulos2002External,Ghisellini2010general}. Second, a lepto-hadronic jet powered by the accretion disc \citep{Blandford1982hydromagnetic} (or which starts out leptonic and entrains hadronic mass) may accelerate leptons and baryons that contribute in different energy bands via different mechanisms \citep[][]{mannheim1993proton,rachen1993extragalactic,MUCKE2003protonBLLac,boettcher2013leptohadronic,Liodakis2020}. 

Recent GeV observations of the high-mass BHXBs Cygnus~X--3 \citep{Tavani_2009CygnusX3} and Cygnus~X--1 \citep{Tavani_2009CygnusX3,malyshev2013high,zanin2016detection}, and TeV observations of SS~433 \citep{abeysekara2018very} suggest that some Galactic jets can accelerate particles to high energy. However, it is not known whether all BHXBs, especially the more abundant population of low-mass BHXBs can routinely produce \gr s. Until now, only the high-mass BHXBs that are characterised by the presence of a strong stellar wind that interacts with the jet have been detected in the GeV and TeV bands \citep[see e.g.,][]{bodaghee2013gamma}.It is thus important to investigate whether the far more populous low-mass BHXBs can also produce \gr s.  In this paper we approach this question by studying the `canonical' low-mass BHXB source \gx , extending our previous work on the `canonical' high-mass BHXB Cygnus~X--1 \citep{kantzas2020cyg}. Similar to AGN jets, the emitting mechanism responsible for any \gr s remains unclear, with both leptonic and hadronic processes considered feasible. We are also interested in exploring how the different composition scenarios may affect the jet dynamics and the interpretation of the jet properties.

In this work, we employ a multi-zone jet model to study the hadronic interactions within the jets, as well as the effect on the dynamics and the electromagnetic signature of low-mass BHXB jets. We examine the bright outburst of \gx\ in 2010 to model the radio-to-X-ray spectrum with the goal of predicting the TeV radiation originating in the jets. Using the case of \gx\  as a model, we assess the likelihood of other, closer low-mass BHXBs to be potential sources for the next generation \gr \ facilities, particularly the Cherenkov Telescope Array (CTA). Such TeV emission may be the signature of efficient CR acceleration inside the BHXB jets, and hence the entire Galactic population of BHXB jets may contribute to the Galactic CR spectrum.

In Section~\ref{sec: gx339} we discuss the physical properties of \gx\ and its spectral behaviour. In Section~\ref{sec: model} we describe the model we use to study the spectrum of \gx. We present our results in Section~\ref{sec: results}, discuss their implication in Section~\ref{sec: discussion} and come to our final conclusions in Section~\ref{sec: summary}. 

\section{\gx}\label{sec: gx339}
\gx\ is a 'canonical' low-mass BHXB discovered in 1973 \citep{Markert1973discovery}. It undergoes outbursts every two-to-three years that last from a few weeks to months \citep{Belloni_1999,Corbel2002NIR,Corbel2003radio,Zdziarski2004distance,Homan_2005outburst,belloni2006,motta2009,Corbel2013universal}. During outbursts, \gx\ rises out of quiescence and launches compact jets that contribute to the radio-to-optical spectrum as the source continues into the hard state \citep{fender2001powerful,fender2004unified,corbel2000coupling,Corbel2003radio,Corbel2013universal,Corbel2002NIR,casella2010fast,Homan_2005outburst,Gandhi2011synchbreak}. Such consistent, repetitive behavior along with extensive and often simultaneous multiwavelength monitoring makes \gx\ a perfect target to better understand the properties of relativistic jets.

Although \gx\ is a well-studied source, its physical parameters are not well constrained because of the weakness of its companion star. Based on optical photometry the orbital period is estimated to be between 14.8 hours and 16.8 hours \citep[][respectively]{callanan199214,cowley2002optical}. The inclination angle is still unknown but is constrained to $< 60$ degrees because of the lack of eclipsing \citep{cowley2002optical}, and the lack of a detection of the companion star means the mass of the black hole is also uncertain. Various current estimates put the mass (in M$_\odot$) between $4-16$ \citep{shidatsu2011suzaku}, 5.8 $\pm$ 0.8 for an orbital period of 1.75 days \citep{hynes2003dynamical}, $>7$ \citep{Munoz2008mass} or 9.8 for a mass function of 1.91 $\pm$ 0.08 M$_\odot$ \citep{heida2017mass}. We adopt the most recent value of $M_{\rm{bh}} = 9.8\, \rm{M_{\odot}}$ of \cite{heida2017mass}. \cite{hynes2004distance} set the distance of \gx\ higher than 6\,kpc, and  \cite{Zdziarski2004distance} derived a value of 8\,kpc while \cite{parker2016nustar} found a distance of $8\pm0.9\,$kpc, which is the distance we adopt here. 

\subsection{Observational constraints in the hard state}
\gx\ has been detected in the optical bands, but the origin of this emission is still not clear. \cite{Tetarenko2020} recently studied its multiwavelength emission and concluded that the optical emission in bright outbursts like the one of 2010 cannot originate exclusively from irradiation of the accretion disc, because unreasonable amounts of energy would be required. Thermal synchrotron emission from the base of the jets could then be considered a good candidate for the optical emission. On the other hand, \gx\ shows a flat spectrum in the radio with a spectral break in the IR band that corresponds to the transition of optically thick to optically thin synchrotron emission \citep{Corbel2002NIR,Gandhi2011synchbreak}. Extrapolating the optically thin IR emission to the X-ray band, significantly underpredicts the optical flux \citep{maitra2009constraining,Gandhi2011synchbreak,tetarenko2019radio}. Hence, if the optical emission originates in the jets, it must come from a different region compared to the IR \citep{markoff2003exploring,corbel2013formation}.

Reflection features, including a broad iron emission line, are also evident in the X-ray spectrum of \gx\ \citep{Nowak2002coronal,garcia2015x,Furst2015complex,parker2016nustar,garcia20192017failed,dzielak2019comparison}. A jet synchrotron component that is beamed perpendicularly away from the accretion disc is very unlikely to produce significant relativistic reflection \citep{Markoff_2004,Reig2021illumination}. 
Furthermore, \cite{uttley2011connection} studied the energy-dependent time lags and found that the instabilities in the accretion disc may be responsible for driving the continuum variability on short and longer-than-second timescales. The large time-lags are due to the travel-time between the illuminating region and the disc where the X-rays are reprocessed, and can be only tens of gravitational radii at most. That indicates that the X-ray continuum should be governed by a single component, and a thermal corona close to the black hole could sufficiently explain it \citep[but also see][for a two-component corona]{Mahmoud2019}.  

Based on these results, we approach the modelling assuming the most conservative case for the jet power: that the radio through IR up to the break is self-absorbed synchrotron from the extended jets, the optical emission is synchrotron emission from thermal particles at the base of the jets, and that the X-ray reflecting power-law is from a separate coronal region.

\subsection{Observational data}\label{sec: observational data}
In this work we use archival quasi-simultaneous data to model the multiwavelength spectrum of \gx\ from radio to X-rays during the hard state of the 2010 outburst. We use the radio data obtained by the Australian Telescope Compact Array (ATCA) on MJD 55263 \citep{Corbel2013universal}, IR data obtained by the Wide-field Infrared Survey Explorer (WISE) on MJD 55266 \citep{Gandhi2011synchbreak}, optical data obtained by the Small \& Moderate Aperture Research Telescope System (SMARTS) on MJD 55263, and X-rays from the \textit{Neil Gehrels Swift Observatory}/X-ray Telescope (\textit{Swift}/XRT) on MJD 55262 \citep{Corbel2013universal} and \textit{Rossi X-ray Timing Explorer}/Proportional Counter Array (\textit{RXTE}/PCA) on MJD 55263 
\citep{Corbel2013universal}. We use the 0.5--4.0\,keV XRT and the 3--45\,keV PCA X-ray data. The IR data are not simultaneous and were obtained 3 days later, but we use them because they show a spectral break crucial for our interpretation of the whole spectrum (see below). There was no significant variability in this time, hence this is a decent assumption to combine these data \citep{Corbel2013universal,corbel2013formation,Connors2019combining}. We also use the upper-limits in the GeV band set by the \textit{Fermi}/Large Area Telescope (LAT) \gr\ telescope during the 2010 outburst to further constrain the highest energy regime of the spectrum \citep{bodaghee2013gamma}. We provide the energy/frequency ranges and the corresponding flux density of all the data we use in Table~\ref{table: data}.

\section{Modelling}\label{sec: model}

In this section we briefly discuss our model, focusing on the interpretation of the free parameters we fit for. A more detailed description of the model can be found in \cite{kantzas2020cyg} and in \cite{lucchini2022BHJet}.

\subsection{Jet properties}\label{sec: jet quantities}
We assume that two compact jets are launched by the accreting black hole with jet base radius $R_0$. The power injected into the jets in the comoving frame $L_{\rm{jet}}$ defines the number density of the cold (non-relativistic) protons in the plasma at the base of the jets as,
\begin{equation}\label{eq: number density jet base}
 n_{0} = \frac{L_{\rm{jet}}}{2\beta_{0,s}\Gamma_{0,s}c\,\pi R_0^2{\rm{(m_pc^2+\langle \gamma_e\rangle m_ec^2(1+1/\beta))}}}, 
\end{equation}
where $\beta_{0,s}\Gamma_{0,s}c$ is the comoving velocity of the plasma in the jet base assumed to be equal to the speed of sound in a relativistic fluid \citep{falcke1995jet,Markoff_2008,crumley2017symbiosis,lucchini2022BHJet}, $\beta= {U_{\rm{e}}}/{U_{\rm{B}}}$ is the plasma beta where $U_{\rm e}$ is the energy density of the electrons and $U_{\rm B}$ is the magnetic field energy density.
For simplicity, we assume equal number density of electrons and protons, but we discuss the implication of this assumption in Section~\ref{sec: discussion}. We further assume that the electron population at the jet base is injected in a thermal Maxwell-J\"{u}ttner (MJ) distribution with a peak-energy of $2.23\,k_{\rm{B}}T_{\rm{e}}$. 

We vary the plasma beta at the jet base to define the strength of the magnetic field, which scales inversely with distance along the jet $z$. Assuming the electron enthalpy is not significant, we define the magnetisation of the jet as
\begin{equation}\label{eq: magnetisation}
    \sigma = \frac{B_0^2}{4\pi n_0{\rm{m_pc^2}}},
\end{equation}
where $B_0$ is the strength of the magnetic field at the jet base.
We do not consider any particular magnetic field configuration (toroidal or poloidal) but merely describe the magnetic field by its total strength $B$.

\subsection{Particle acceleration}\label{sec: particle acceleration}
At some distance $z_{\rm{diss}}$ along the jet axis, energy is dissipated into accelerating a fraction of the thermal particles into a non-thermal power-law. We assume that the accelerated particles carry a fixed fraction of the jet power, and in particular, we conservatively fix the power of the non-thermal leptons to be 0.02$L_{\rm jet}$ and of the protons to be 0.05$L_{\rm jet}$.

We allow $z_{\rm{diss}}$ to vary as a fitted parameter and, for the case of the leptonic populations, we assume constant re-acceleration along the jet, but we constrain the proton acceleration to occur only between $z_{\rm diss}$ and $10\,z_{\rm diss}$ in order to limit the required power. Because the most compact part of the jet produces the non-thermal particles, this dissipation region also corresponds to the region where the synchrotron radiation breaks from flat/inverted due to self-absorption, to steep/optically thin. After predictions by \citet{markoff2001jet}, \citet{Corbel2002NIR} confirmed that this break typically falls in the NIR band during hard states, and we chose the epoch here because of high-quality observations by \cite{Gandhi2011synchbreak} that could pinpoint the synchrotron break frequency to be $\rm{4.6^{+3.5}_{-2.0}\times 10^{13}\,Hz}$. To match this frequency, we fix the particle acceleration region at 2600\, $r_{\rm{g}}$ from the black hole \citep[see also][]{Connors2019combining}.

The accelerated particles follow a power-law in energy of the form
\begin{equation}\label{eq: injection term}
  {\rm{d}}n\left(E\right)\propto E^{-p}\times \exp{\left(-E/E_{\rm{max}}\right)}.
\end{equation}
In principle, the power-law index $p$ depends on the acceleration mechanism and may differ between electrons and protons, but we choose to use the same for both populations for simplicity. 

In equation~\ref{eq: injection term}, $E_{{\rm{max}}}$ is the maximum particle energy constrained by energy losses and/or escape. In this work, the maximum electron energy is limited by synchrotron losses and the maximum proton energy is limited by the lateral escape from the jet region. The maximum attainable energy is self-consistently calculated along the jet by equating the characteristic timescales of the losses to the acceleration timescale. The characteristic acceleration timescale $t_{\rm{acc}} = 4E/(3f_{\rm{sc}} \rm{ec}B)$ depends on the acceleration efficiency parameter $f_{\rm{sc}}$ that we take to be close to maximum, namely $f_{\rm{sc}}=0.01$ \citep{jokipii1987rate,aharonian2004very}. We plot the characteristic timescales versus the particle kinetic energy for the population of the accelerated protons in appendix~\ref{appendix: proton timescales}.

The fractional number of accelerated particles with respect to the total number of particles $f_{\rm{nth}}$ depends on the acceleration mechanism as well. This number may not be constant along the jet. We parametrize the density of the accelerated particles following:
\begin{equation}\label{eq: pldist}
    n_{\rm{nth}} = n_{\rm{th}}\, f_{\rm{nth}} \, \left( \frac{\log_{10}(z_{\rm{diss}})}{\log_{10}(z)} \right)^{f_{\rm{pl}}},
\end{equation}
where $f_{\rm{pl}}>0$ is a free parameter accounting for our ignorance about the exact nature of the dissipation. ($n_{\rm{nth}}$) $n_{\rm{th}}$ is the number density of the (non-)thermal particles. The physical motivation behind such an assumption is the fact that it leads to the characteristic inverted spectrum between radio and optical wavelengths detected in BHXBs (see discussion in \citealt{Lucchini2021correlation}).

The minimum energy of the accelerated particles depends on the injected distributions in the base. We assume that the minimum energy for accelerated protons is the rest mass energy ($\rm{m_pc^2}$). This choice is intended purely to limit the number of free parameters; we discuss its implication below. We take the peak of the MJ distribution $2.23k_{\rm{B}}T_{\rm{e}}$  to be the minimum energy of the accelerated electrons. We further define a heating parameter $f_{\rm{heat}}$
\begin{equation}\label{eq: heat}
    E_{\rm{e,min}} = 2.23f_{\rm{heat}}k_{\rm{B}}T_{\rm{e}}.
\end{equation}
The physical motivation behind this assumption is that along with the electron acceleration, some extra heating has been reported by numerical simulations \citep{Sironi2009,Gedalin_2012,Plotnikov2013,Sironi2013,Sironi2014, Melzani2014, crumley2017symbiosis}. The value of this parameter is not well constrained, but we set it to be $f_{\rm heat}$< 10  \citep{Sironi2009,Sironi2011,crumley2019kinetic}.

\subsection{Radiative Processes}\label{section: radiative processes}
\subsubsection{Leptonic processes}
Following \cite{kantzas2020cyg}, the leptonic radiative processes we take into account are cyclo-synchrotron radiation and inverse Compton scattering (ICS), where the cyclo-synchrotron photons are further upscattered via the synchrotron-self Compton mechanism (SSC) along the jets. Further photon targets for the ICS are the disc photons. We also take into account a precise treatment of pair production due to photon annihilation and pair annihilation to electron-positron pairs \citep{coppi1990reaction,bottcher1997pair}.

\subsubsection{Hadronic Processes}\label{hadronic processes}
Accelerated protons interact with the bulk cold protons of the jet and, via proton-proton (pp) interactions, lead to pion production. Neutral pions decay into \gr s and charged pions into secondary electrons and neutrinos via the muon decay channel \citep{mannheim1994interactions}. Photomeson interactions between the accelerated protons and target photons (p\g) lead to similar distributions of secondary particles. The target photons we consider here are: the thermal radiation of the accretion disc and the non-thermal radiation originating in the compact jet. Finally, we also account for photopair interactions that lead to the formation of pairs, after the inelastic collision between protons and photons. We use the semi-analytical formalism of \cite{kelner2006energy} and \cite{kelner2008energy} for pp and p\g\ interactions, respectively. For the full description of the treatment of the cascades, see \cite{kantzas2020cyg}. For the case of \gx , no photon field is significant enough to attenuate the GeV and TeV emission (also see the discussion below).

\subsection{Accretion disc and thermal corona}
We assume a standard geometrically thin, optically thick accretion disc truncated at some innermost radius $R_{\rm{in}}$ with temperature $T_{\rm{in}}$ \citep{shakura1973black,frank2002accretion}. We describe the disc luminosity $L_{d}$ in terms of  Eddington luminosity $L_{\rm{Edd}} = 4\pi {\rm{G}}M_{\rm{bh}} {\rm{m_pc}}/\sigma_{\rm{T}}$. We further assume the existence of a hot electron plasma of temperature $T_{\rm{cor}}$, in a spherical region centered on the black hole, normalized by a radius $R_{\rm{cor}}$, and of optical depth $\tau_{\rm{cor}}=n_{\rm e}R_{\rm cor}\sigma_{\rm T}$. These hot electrons upscatter the disc photons to higher energies. We require the existence of such a plasma to be able to model both the X-ray spectrum and properly account for the measured hard timing lags as mentioned in Section~\ref{sec:intro} \citep[and see e.g.,][and discussion below]{Connors2019combining}. 

\begin{table*}
	\begin{center}
		\setlength{\tabcolsep}{6pt} 
		\renewcommand{\arraystretch}{1.2} 
		\begin{tabular}[b]{lcccc}\hline\hline
			Observatory & log Frequency (Hz) & log Energy (eV) & Flux Density (mJy$^{\rm{a}}$)& Reference\\ \hline
		    ATCA& $\begin{array}{cc}
			      9.74 & 9.94
			\end{array}$&
			$\begin{array}{cc}
			      -4.64 & -4.44
			\end{array}$
		     &$\begin{array}{cc}
			      10.2\pm 0.1 & 11.3 \pm 0.1
			\end{array}$&\citealt{Corbel2013universal}\\
			WISE& $\begin{array}{cc}
			      13.13 & 13.41\\13.81  & 13.95
			\end{array}$& $\begin{array}{cc}
			      -1.25 & -0.97\\-0.57  & -0.43
			\end{array}$ &$\begin{array}{cc}
			      87\pm 8 & 80\pm 7\\64 \pm 5  & 55\pm 4
			\end{array}$& \citealt{Gandhi2011synchbreak}\\
			SMARTS& $\begin{array}{cc}
			      14.25 & 14.40\\14.57  & 14.73
			\end{array}$& $\begin{array}{cc}
			      -0.13 & 0.01\\0.18  & 0.35
			\end{array}$ &$\begin{array}{cc}
			      47\pm 5 & 50\pm 5\\54 \pm 5  & 92\pm 29
			\end{array}$& \citealt{Buxton_2012}\\
			\textit{SWIFT}/RXT &17.08--18.0&2.7--3.7& $0.2$ at $3\,$keV & \citealt{Corbel2013universal} \\
			\textit{RXTE}/PCA &17.9--18.9&3.5--4.5& $0.2$ at $3\,$keV & \citealt{Corbel2013universal} \\
			\hline
		\end{tabular} 
		\caption{ The observational multiwavelength data we use in this work.\newline  $^{\rm{a}}$mJy$\,= 10^{-26}\,\rm{erg\,cm^{-2}\,s^{-1}\,Hz^{-1}}$
		\label{table: data}}
	\end{center}
\end{table*}

\begin{table}
	\begin{center}
		\setlength{\tabcolsep}{3pt} 
		\renewcommand{\arraystretch}{1.2} 
		\begin{tabular}{lrl}\hline
			parameter & value & description \\
			\hline
			$M_{\rm{BH}}\, \left(\rm{M_{\odot}}\right)$& 9.8      & mass of the black hole$^{\dagger}$\\
			$\theta_{\rm{incl}}$ 		& 40$^\circ$        & inclination angle$^{\dagger}$\\
		$D\, \rm{\left(kpc\right)}$ 				& 8              & distance of the source$^{\star}$\\
			$h=z_0/R_0$ 			& 2                 & initial jet height to radius ratio\\
			$z_{\rm{diss}}\, \left(r_{\rm g}\right)$ 		& $2600$            & particle acceleration region$^{\ast}$\\
			$z_{\rm{diss,max}}$& $10z_{\rm diss}$            & maximum proton acceleration region\\			
			$z_{\rm{max}}\, \left(r_{\rm g}\right)$ 		& $10^8$            & maximum jet height\\
            $f_{\rm sc}$ & 0.01 & particle acceleration efficiency parameter\\
            $P_{\rm e}/L_{\rm jet}$ & 0.02 & power of non-thermal electrons\\
            $P_{\rm p}/L_{\rm jet}$& 0.05& power of non-thermal protons\\
		$R_{\rm{in,disc}}$ 	& $\rm{R_0}$          & disc innermost radius (see Table \ref{table: parameters for models})\\
		$R_{\rm{out,disc}}\, \left(r_{\rm g}\right)$ 	& $10^{5}$          & disc outermost radius\\
	    $\rm{N_{H}\,(10^{22}\,cm^{-2})}$ & $0.6$ & absorption coefficient$^{\ddagger}$\\
	    \textit{refl} & 0.29 & reflection fraction$^{\diamond}$ \\
			\hline
		\end{tabular} 
		\caption{The fixed parameters of our models, see text for further discussion.  $^{\dagger}$\protect\cite{heida2017mass},
		$^{\star}$\protect\cite{parker2016nustar},
		$^{\ast}$\protect\cite{Gandhi2011synchbreak},
		$^{\ddagger}$\protect\cite{garcia20192017failed},
		$^{\diamond}$\protect\cite{Magdziarz1995}
		} 
		\label{table:common parameters}
	\end{center}
\end{table}

\section{Results}\label{sec: results}
In this section, we present the results for the best fits of our model to the multiwavelength spectrum of \gx . We explore three different model scenarios: one purely leptonic, and two lepto-hadronic models. For the purely leptonic model, we assume that the non-thermal electrons follow a power-law with $p=2.2$ \citep[][]{Corbel2002NIR,Gandhi2011synchbreak}. For the two hadronic models, we explore both a soft ($p=2.2$) and a hard ($p=1.7$) particle power-law, respectively. For all models we fix some common parameters as shown in Table~\ref{table:common parameters}. We choose the ratio between the height of the jet base and its radius to be constant and equal to 2 \citep{Maitra_2011,crumley2017symbiosis}. The maximum height of the jet is fixed at a large enough value, so it does not influence the spectrum in the radio band via the self-absorption cutoff, and we choose the maximum reasonable particle acceleration efficiency parameter $f_{\rm{sc}}=0.1$, which results in maximum proton energies of the order of tens of TeV in the hadronic models. We tie the truncation radius of the thin accretion disc to the jet base radius to reduce model degeneracy because the disc does not contribute to the electromagnetic spectrum at all. We use the \texttt{tbabs} model to account for the neutral photoelectric absorption in the intergalactic medium, using the cross-sections by \cite{Verner1996} and the cosmic abundances by \cite{Wilms2000}, where the absorption coefficient $N_{\rm{H}}$ sets the X-ray absorption column. We use the non-relativistic \texttt{reflect} function to treat in a simplified way the reflection detected in \gx, parametrised primarily via the reflection fraction $\textit{refl} =\Omega /2\pi$, which indicates the amplitude of the reflected spectrum \citep{Magdziarz1995}. We choose this simple model in order to minimize the free parameters used to describe the X-ray spectrum, which is well-fit by a power-law. Our focus is on constraining the jet physics that drives the \gr\ band, thus we retain most of the free parameters for that model.

We use the Interactive Spectral Interpretation System (\texttt{ISIS}; \citealt{houck2000ISIS}) to forward fold the model into X-ray detector space, and to find the statistical best fit to the data presented in Section~\ref{sec: observational data}. We use the \texttt{emcee} function to explore the parameter space using a Markov Chain Monte Carlo (MCMC) method \citep{Foreman_Mackey_2013}. We initiate 20 walkers per free parameter and perform $10^4$ loops. We reject the first 50\% of the run as the “burn-in'' period. We provide the $1\,\sigma$ uncertainties in Table~\ref{table: parameters for models}, along with the results of the best fit for each model. 

\begin{figure}
    \includegraphics[width=1\columnwidth]{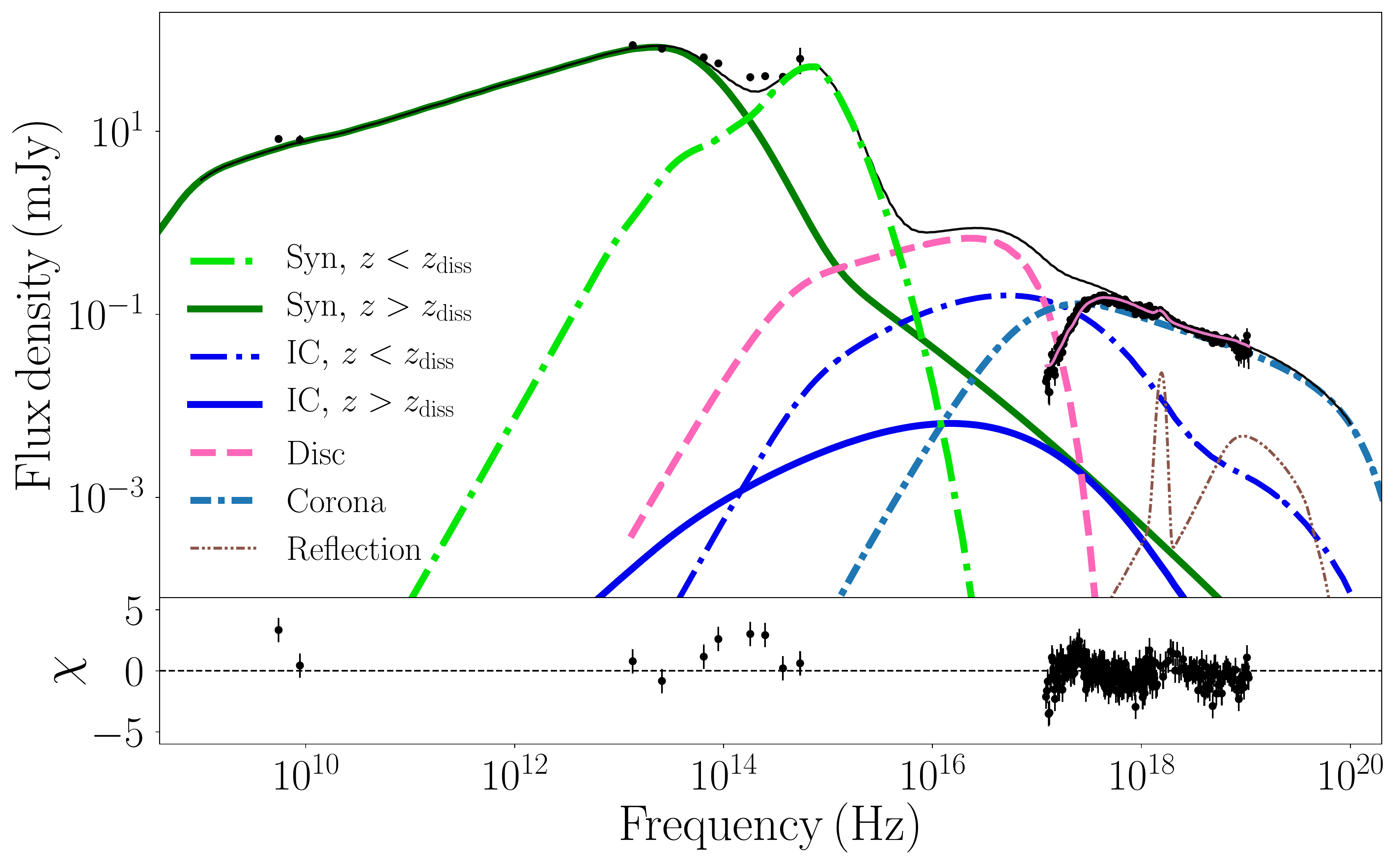}
    \caption{The best fit with the $\chi$-residuals of the multiwavelength spectrum of the 2010 outburst of \gx\ assuming a purely leptonic model. The solid black line shows the total intrinsic emission, the red line shows the X-ray absorbed emission, and the rest of the components are explained in the legend. }
    \label{fig: leptonic}
    \includegraphics[width=1\columnwidth]{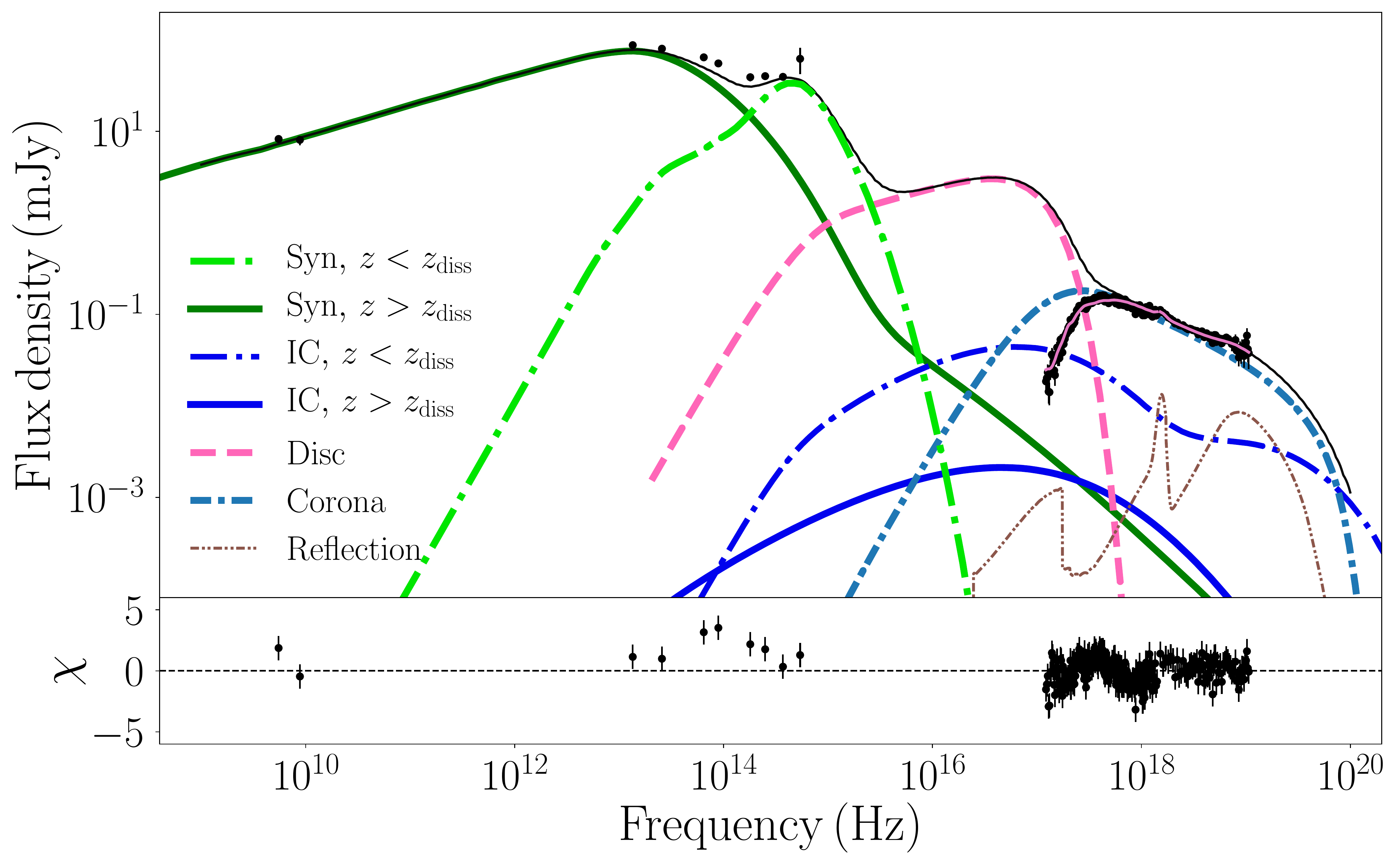}
    \caption{Similar to Fig.~\ref{fig: leptonic} but for the lepto-hadronic model with $p_{\rm{e}}$= 2.2 and $p_{\rm{p}}$= 2.2 power-law index. }
    \label{fig: hadronic e2.2/p2.2}
    \includegraphics[width=1\columnwidth]{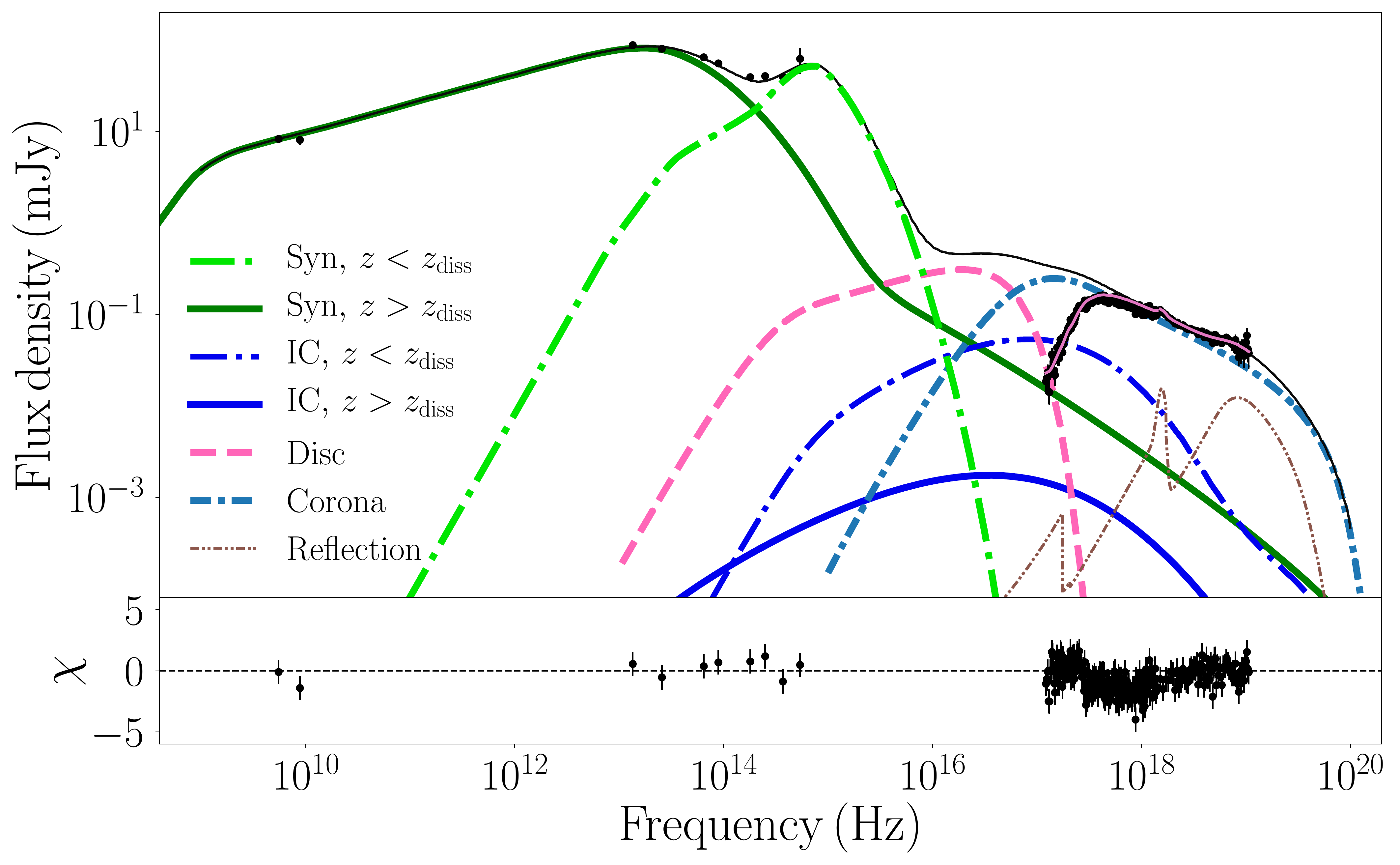}
    \caption{Similar to Fig.~\ref{fig: leptonic} but for the lepto-hadronic model with $p_{\rm{e}}$= 1.7 and $p_{\rm{p}}$= 1.7 power-law index. }
    \label{fig: hadronic e1.7/p1.7}    
\end{figure}

In Figures~\ref{fig: leptonic}--\ref{fig: hadronic e1.7/p1.7} we show the best fits of the multi-wavelength spectrum of \gx\ for the three different models we explore. In Fig.~\ref{fig: leptonic}, we show the purely leptonic model, whereas in Figures~\ref{fig: hadronic e2.2/p2.2} and \ref{fig: hadronic e1.7/p1.7} the results of the lepto-hadronic models.

\begin{table}
		\setlength{\tabcolsep}{5pt} 
		\renewcommand{\arraystretch}{1.5} 
		\begin{tabular}[b]{lccc}\hline
			parameter \textbackslash model & leptonic   & hadronic soft & hadronic hard \\
			$p_{\rm{e}}$    & 		2.2&2.2& 1.7 \\
			$p_{\rm{p}}$    & 		-&2.2& 1.7 \\
			\hline
		$L_{\rm{jet}} \begin{array}{c}
			      \left(10^{-3}\,L_{\rm{Edd}}\right)  \\ 
			      \rm{\left(\times 10^{36}\,erg\,s^{-1}\right)}
			\end{array}$ 		 & $ \begin{array}{c} 2.5_{-2}^{+5} \\ 3_{-2}^{+6} \end{array} $ & $ \begin{array}{c} 70_{}^{+100} \\ 90_{-83}^{+117}\end{array} $	& $ \begin{array}{c} 50_{}^{+60} \\ 70_{-60}^{+70}\end{array} $
		    \\ 
		$R_0 \left(r_{\rm{g}}\right)$ & $ 100_{}^{+100}$    & $     110_{}^{+100}$ & $90_{}^{+90}$ \\
        $T_{\rm{e}}\,(\rm{keV})$ &$1600_{-600}^{+2400}$& $2100_{-2000}^{+2300}$ &$2000_{-1900}^{+2000}$ \\
			$f_{\rm{pl}}$         & $4_{-3}^{+5}$ & $4_{-3}^{+5}$& $4_{-3}^{+5}$ \\
			$f_{\rm{heat}}$         & $5_{-5}^{+7} $ & $ 8_{-6}^{+9}$    & $ 7_{-6}^{+9}$ \\
			$\beta$         & $0.2_{-0.1}^{+1.5}$ & $ 0.2_{-0.1}^{+0.4}$ & $0.04_{}^{+0.04}$ \\
			$L_d\,(10^{-3}\,L_{\rm Edd})$ 	& $5_{-4}^{+9}$ &$ 23_{-22}^{+24}$ & $2_{}^{+3}$ \\
			$T_{\rm{cor}}\, \rm{\left(keV\right)}$ 		    & $170_{-100}^{+200}$ & $ 55_{}^{+350}$       & $60_{-50}^{+70}$  \\
			$R_{\rm{cor}}\, \left(r_{\rm{g}}\right)$ 		    &$300_{-200}^{+300}$ & $160_{-155}^{+460}$  & $460_{-440}^{+470}$  \\
			$\tau_{\rm{cor}}$ 			        & $0.6_{-0.4}^{+0.6}$ &$ 0.7_{-0.6}^{+0.8}$    &$0.7_{}^{+0.7}$ \\
			\hline \hline
			$\chi^2$/DoF &250/233&240.8/233&190/233\\
			\hline \hline
			$\sigma$    & 		1.7&0.03& 0.1 \\
			$B_0\,\left(\rm{G}\right)$ 			&$2\times 10^5$   & $1\times 10^6$    &$2\times 10^{5}$ \\
			$B\,\left(\rm{G}\right)@\, z_{\rm{diss}}$ 	&$1\times 10^4$  &$6\times 10^3$     & $1\times 10^{4}$\\
			$E_{\rm{p,max}}\,\rm{\left(eV\right)}$ 	& - & $2.8\times 10^{13}$    & $2.7\times 10^{13}$ \\
			$E_{\rm{e,max}}\,\rm{\left(eV\right)}$ 	&$1\times 10^{8}$&$5.2\times 10^{10}$    & $5.3\times 10^{10}$  \\
			\hline
		\end{tabular} 
		\caption{Parameters for the three fitted models, distinguished via the power-law index of the accelerated electrons $p_{\rm{e}}$ and protons $p_{\rm{p}}$. We show the free parameters and the 1\,$\sigma$  uncertainties as discussed in Section~\ref{sec: model} before the double line. Below the double line are indicative evaluated quantities of the plasma magnetisation, the magnetic field, the total luminosity of the accelerated proton/electron population and the maximum energy of the protons/electrons at the particle acceleration region.\newline
        }
		\label{table: parameters for models}
\end{table}

\begin{figure}
    \centering
    \includegraphics[width=1.\columnwidth]{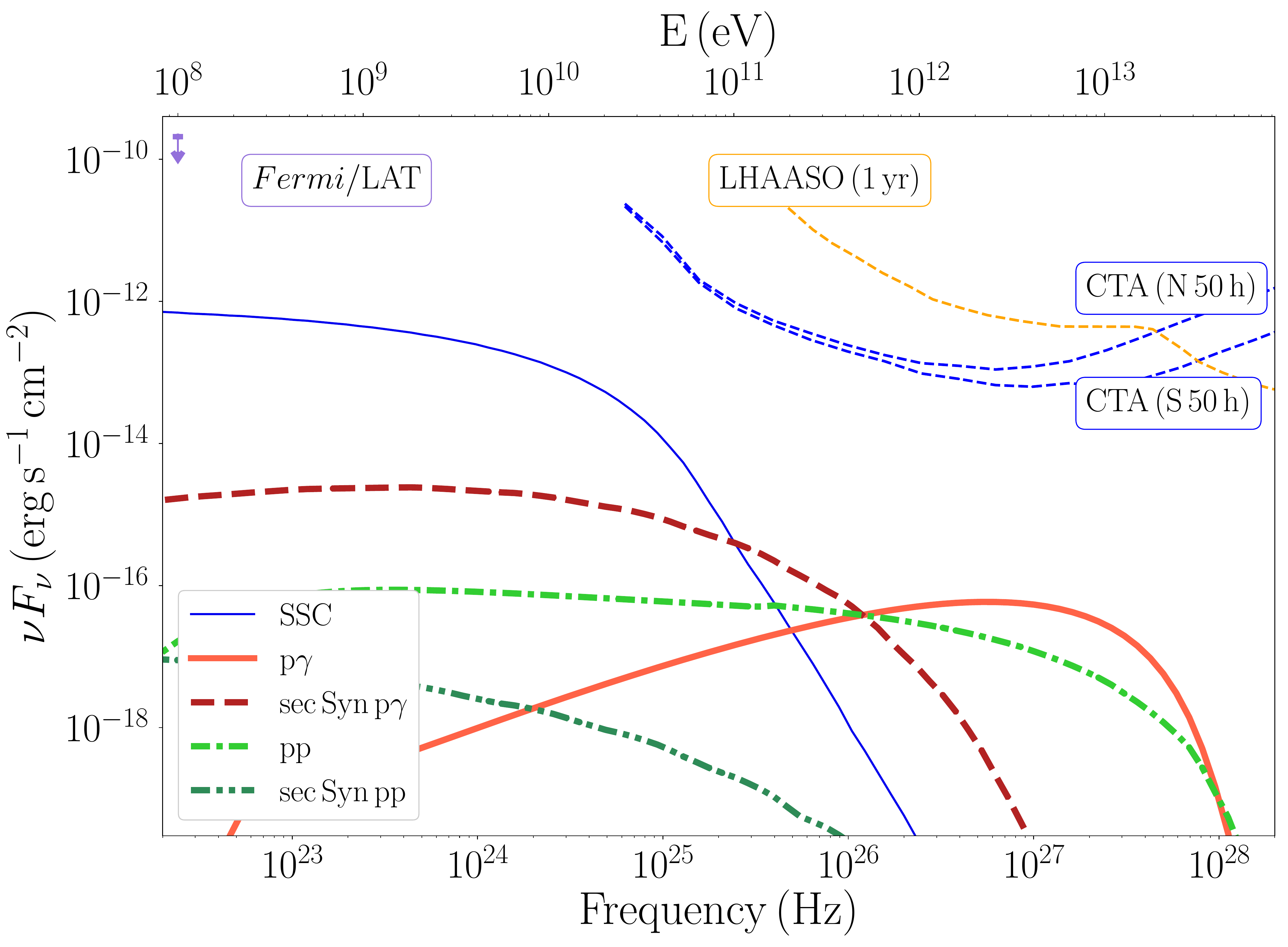}
    \caption{The \gr\ spectrum for the lepto-hadronic model with $p_{\rm{e}}$= 2.2 and $p_{\rm{p}}$= 2.2 power-law index. We compare the predicted spectrum to the \textit{Fermi}/LAT upper limits of the 2010 outburst \citep{bodaghee2013gamma}, the sensitivity of LHAASO after one year of operation \citep{bai2019large}, and the predicted 50-hour sensitivity of the North and South site of CTA (from \href{www.cta-observatory.org}{www.cta-observatory.org}). The solid red line shows the pion bump from p\g, the dashed red line shows the synchrotron radiation from secondary leptons from p\g , the dash-dotted  green line shows the pion bump from pp interactions, and the dash-double dotted line shows the synchrotron radiation from secondary leptons from pp. }
    \label{fig: hadronic e2.2/p2.2 CTA}
    
    \includegraphics[width=1.\columnwidth]{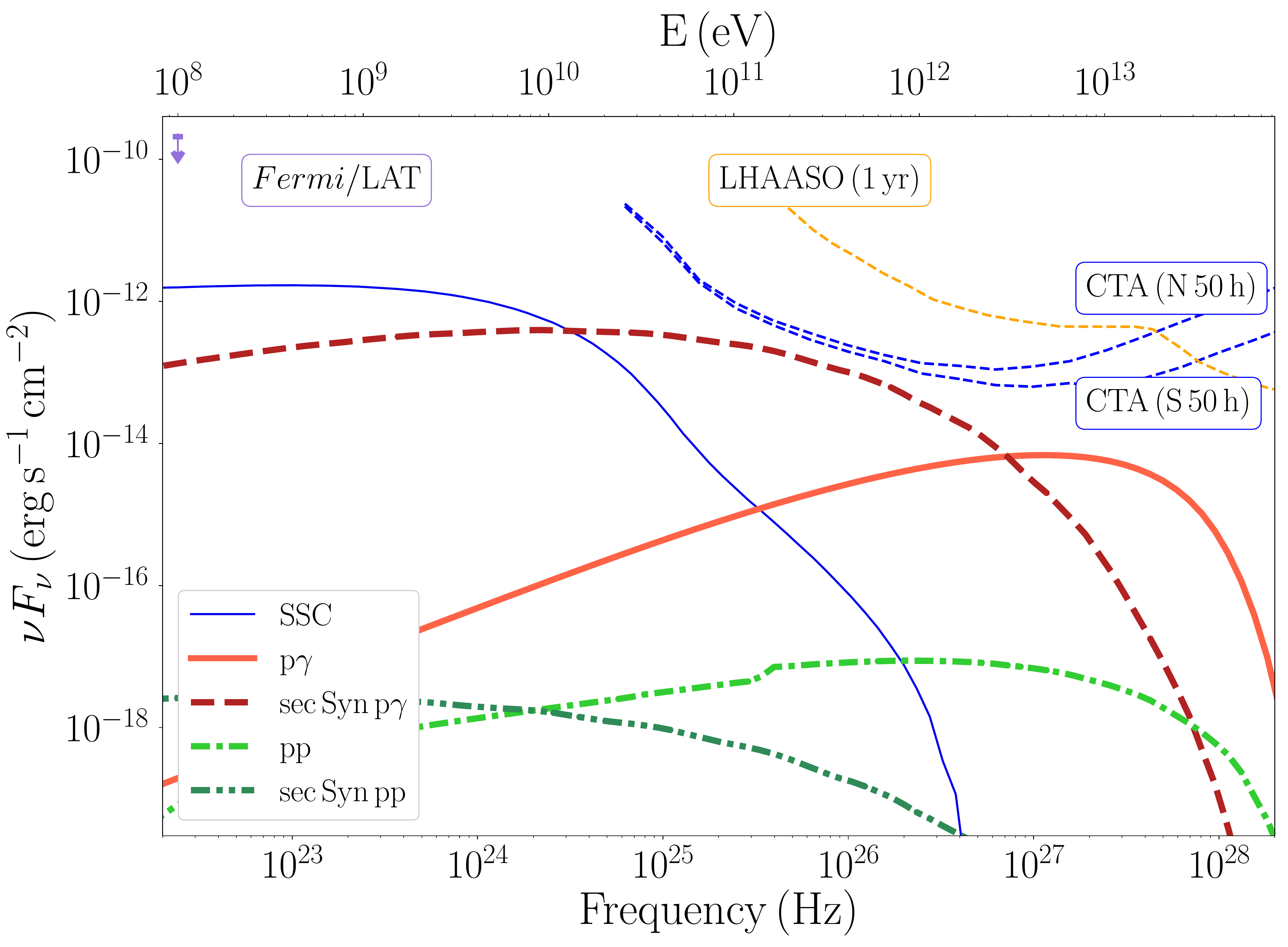}
    \caption{Similar to Fig.~\ref{fig: hadronic e2.2/p2.2 CTA} but for the lepto-hadronic model with $p_{\rm{e}}$= 1.7 and $p_{\rm{p}}$= 1.7 power-law index.}
    \label{fig: hadronic e1.7/p1.7 CTA}
\end{figure}

The unique contribution of the hadronic processes can only be seen in the TeV \gr\ band, because the purely leptonic model cannot produce significant emission at GeV and above. In Figures~\ref{fig: hadronic e2.2/p2.2 CTA} and \ref{fig: hadronic e1.7/p1.7 CTA} we show the predicted GeV to 100\,TeV \gr\ spectrum of \gx . The primary-accelerated electrons dominate in the GeV regime via SSC. The hadronic processes dominate in the TeV energy band, in particular, the neutral pion decay from both pp and p\g \ collisions as well as the synchrotron radiation of secondary pairs from the latter. Because we set the acceleration efficiency parameter $f_{\rm{sc}}$ to a high value, the protons are able to achieve high energies of the order of $\sim 10^{13}\,$eV, producing \gr s of the order of TeV.

\section{Discussion}\label{sec: discussion}

\subsection{Multi-wavelength spectrum and jet dynamics}\label{sec: MW spectrum and dynamics}
Our results with our new lepto-hadronic multi-zone jet model confirm earlier results that stratified jets can self-consistently reproduce the radio-to-X-ray spectrum, together with a thin accretion disc including reflection \citep{markoff2001jet,Zhang2010neutrinoLMXRBs,Kylafis2018correlation,Connors2019combining,lucchini2022BHJet}. However, compared to earlier works \citep[e.g.,][]{Markoff2005}, we can also better reproduce the significantly inverted radio-to-IR spectrum by introducing a decreasing particle acceleration efficiency along the jets \citep{Lucchini2021correlation}. We see however in Table~\ref{table: parameters for models} that the parameter $f_{pl}$ controlling this effect cannot be well-constrained by the data, and we can only set an upper-limit.

Apart from particle acceleration, we require significant electron heating of the thermal population\citep{Sironi2009,Gedalin_2012,Plotnikov2013,Sironi2013,Sironi2014, Melzani2014,crumley2019kinetic} to reproduce both the optical and IR bands as jet synchrotron emission. In particular, we find that the scenario where optical emission originates from the jet base and the IR emission originates from the particle acceleration region $z_{\rm diss}$ is consistent with the data. An alternative scenario is that both the IR and the optical emission originate in a hot flow that consists of thermal and non-thermal electrons \citep[][]{Poutanen_2014,Kosenkov2020colors}, a scenario that better describes the soft states \citep[][]{Kosenkov2018superhump}. Further simultaneous IR-to-optical observations in the hard state would be able to test this scenario, as well as simultaneous polarisation measurements across the entire optical/IR band (although see e.g., \citealt{Russell2008poloarized} for measurements prior to the 2010 outburst). 

In both the leptonic and lepto-hadronic scenarios the shape of the radio-to-X-ray spectrum of \gx\ looks identical and the radiative mechanisms are also the same. The spectral shape is determined primarily by the jet geometry and dynamics, which are similar between the scenarios. However, for the case of the lepto-hadronic models, where we assume equal number density of accelerated electrons and protons, we require much more power injected into the jet base than for the purely leptonic model, which is a well-known issue with hadronic models \citep[see e.g.,][]{pepe2015lepto,abeysekara2018very,kantzas2020cyg}.

To fit the optical emission with thermal synchrotron emission from the base of the jets while the accelerated particles fit the radio-to-IR, we require high electron temperature. This radiation leads to a curved IC spectrum in the soft X-rays, so another component is required to explain the hard power-law. If it can be confirmed that the optical emission is jet synchrotron (via polarisation for instance), then the need for a second component to fit the X-rays will be more robust. For this reason we have added a simple thermal corona model, which together with reflection, can well account for the X-ray spectrum, but is otherwise independent of the jet parameters. In reality, these components should be linked, but it is well known that spectral information alone is often not enough to probe the detailed geometry of the corona, which is the case in our work here as well \cite[see e.g.,][]{DelSanto2008variability,Droulans_2010,Reig2015correlation,Reig2021illumination,Kylafis2018correlation,Connors2019combining,cao2021evidence}.

When protons are accelerated, the hadronic interactions contribute with additional flux in the \gr \ regime of the spectrum. For the scenario with a hard proton power-law index of $p_{\rm p}$= 1.7, producing significant TeV flux detectable by CTA would require a non-physical amount of power dissipated into proton acceleration. By constraining the non-thermal proton power to 5\% of the jet power, we see that the TeV flux does not exceed the CTA sensitivity (see Fig.~\ref{fig: hadronic e1.7/p1.7 CTA}). A more typical power-law index of $p_{\rm p}$= 2.2 produces even less GeV and TeV flux. In addition, both of these models require strongly matter-dominated outflows even at their launching point ($\sigma \lesssim 0.1$). Such a low magnetisation raises issues of physicality for these models, since the final bulk Lorentz factor of the flow is expected to be on the order of the initial magnetisation $\sigma$ \citep{Komissarov2007magnetic,Komissarov2009ultrarelativistic,Tchekhovskoy2008simulations,Tchekhovskoy_2009,chatterjee2019accelerating}. Specifically, BHXB jets consistently show at least mildly relativistic velocities of $\Gamma \sim 2-3$ in several systems \citep{mirabel1994superluminal,fender2001powerful,fender2004unified,casella2010fast,Miller-Jones2012first}. Such a low initial magnetisation would struggle to explain the bulk acceleration of the flow unless further energy is available by, e.g., thermal pressure. However, numerical simulations show that a jet 'sheath' forms where the originally Poynting-flux dominated 'spine' interacts and entrains the surrounding disc wind, resulting in a region with much lower magnetisation \citep{McKinney2006,Moscibrodzka2016grmhd,Nakamura_2018,chatterjee2019accelerating}. The instabilities that form along this boundary are expected to be sites of reconnection and particle acceleration \citep{rieger2004shear,Faganello2010collisionless,rieger2019shearing,Sironi2020}. Thus, although our approach is quite simplistic, it would be consistent with the emission occurring along this boundary as suggested by recent radio observations of AGN jets, such as M87 \citep[][]{Hada2016VLBIM87} or Cen~A \citep[][]{Janssen2021ehtCenA}, and GRMHD simulations \citep[e.g.][]{Moscibrodzka2013,Davelaar2018}. Although BHXB jets cannot be resolved by current facilities, similar scenarios may apply to them since the systems are likely to be governed by the same physical laws \citep{Heinz2003dependence,Merloni2003fundamental,Falcke2004unify}.

\subsection{Particle distributions}
\begin{figure}
    \centering
    \includegraphics[width=1.\columnwidth]{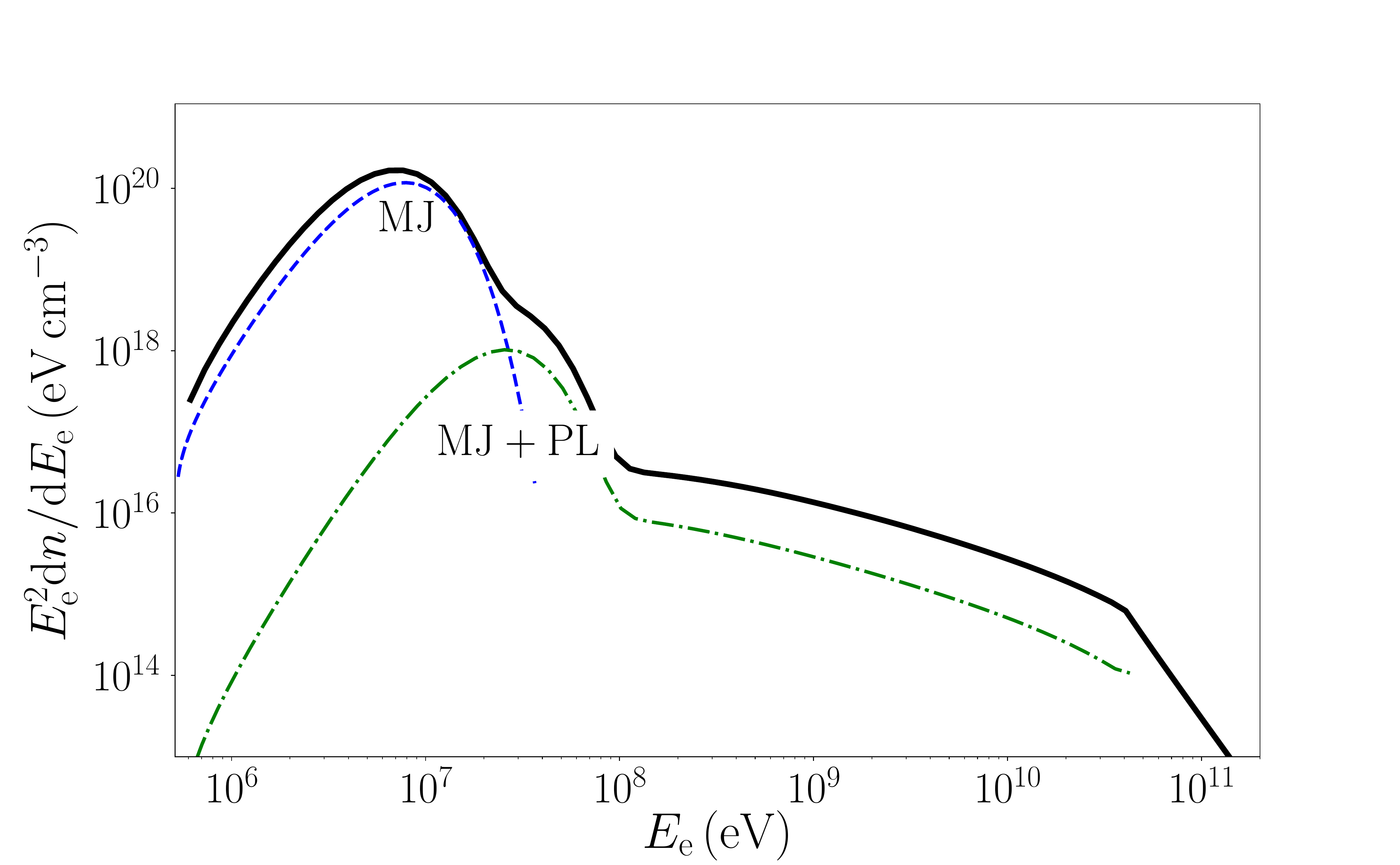}
    \caption{The total electron number density of the jets multiplied by the square of the electron energy for the lepto-hadronic model (solid black line) with $p_{\rm{e}}=1.7$ and $p_{\rm p}$= 1.7 power-law index. We also show the  contribution of the Maxwell-J\"{u}ttner distribution at the jet base (MJ; dashed blue line), and the contribution of the MJ plus the power-law tail of accelerated electrons (MJ+PL; dash-dotted green line) at the particle acceleration region $z_{\rm{diss}}$.}
    \label{fig: hadronic e1.7/p1.7 electrons}

    \includegraphics[width=1.\columnwidth]{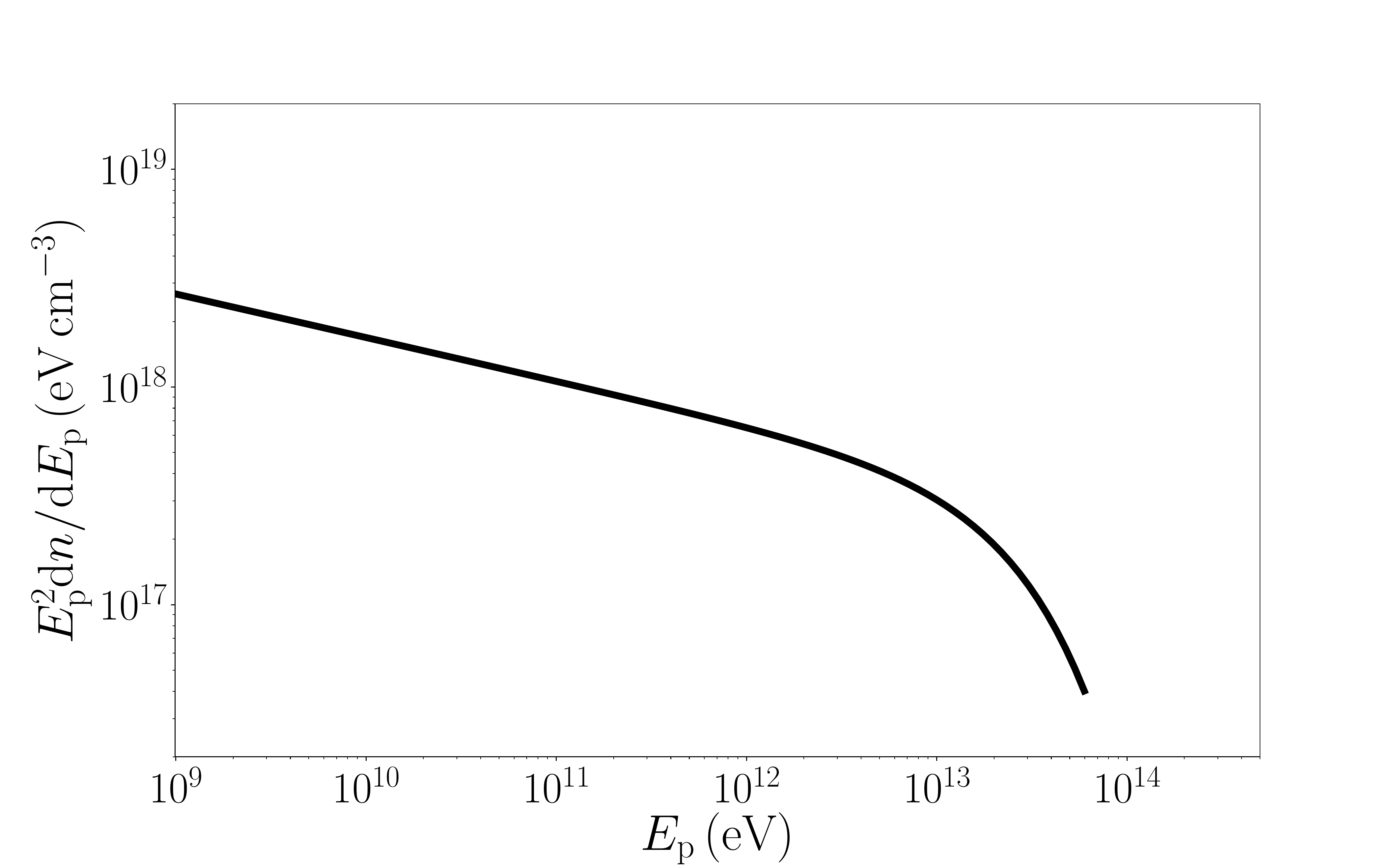}
    \caption{Similar to Fig.~\ref{fig: hadronic e1.7/p1.7 electrons} but for the population of protons for the lepto-hadronic model with $p_{\rm{e}}$= 2.2 and $p_{\rm{p}}$= 2.2 power-law index.}
    \label{fig: hadronic e2.2/p2.2 protons}

    \includegraphics[width=1.\columnwidth]{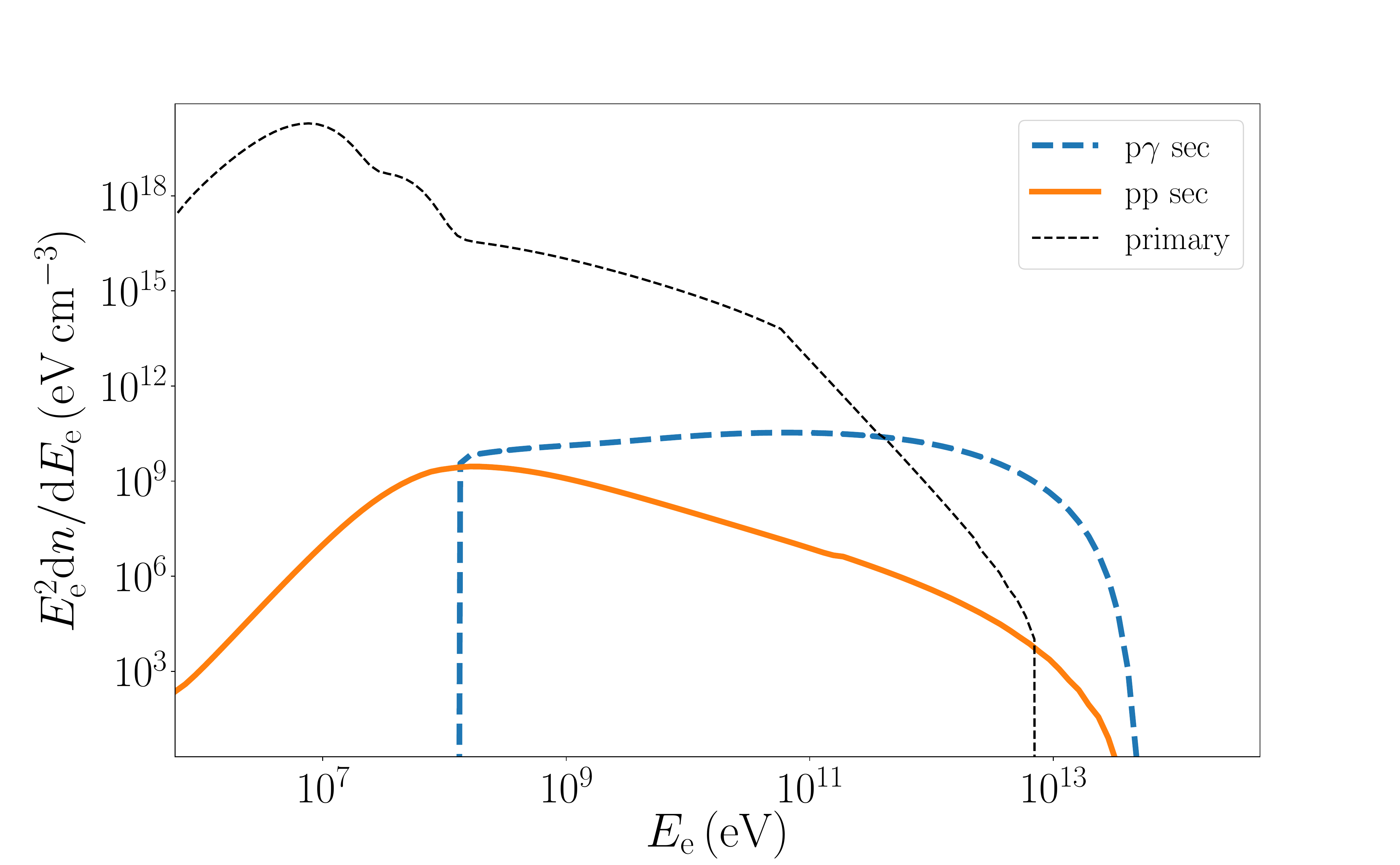}
    \caption{Similar to Fig.~\ref{fig: hadronic e1.7/p1.7 electrons} but for the secondary pairs of the lepto-hadronic model with $p_{\rm{e}}$= 2.2 and $p_{\rm{p}}$= 2.2. We also plot the distribution of the primary electrons for comparison}
    \label{fig: hadronic e2.2/p2.2 secondary pairs from pg}
\end{figure}

In Fig.~\ref{fig: hadronic e1.7/p1.7 electrons} and ~\ref{fig: hadronic e2.2/p2.2 protons} we plot the total distribution of the primary electrons and protons, respectively, integrated along the jets. The MJ-only  distribution at the jet base dominates the lower energy regime, with its peak defined by the free parameter $T_{\rm{e}}$ (see Table~\ref{table: parameters for models}), while the higher energy electrons originate mostly at the first particle acceleration region $z_{\rm diss}$. The shifting of the thermal peak between the two shows the effect of the $f_{\rm{heat}}$ parameter. The fact that the slope is steeper than $p_{\rm e}=2$ indicates that the synchrotron cooling break occurs below $\sim10^9$ eV.  

In Fig.~\ref{fig: hadronic e2.2/p2.2 secondary pairs from pg} we plot the differential number density of the secondary pairs from pp and p\g\ for the lepto-hadronic model with $p_{\rm p}$= 2.2. We also include for comparison the total distribution of the primary pairs of the jets. We note that the secondary pairs from p\g \ are synchrotron cooled and hence their spectrum is flat. The excess of particles around $\sim 10^{12}\, \rm eV$ is responsible for the TeV flux of Fig.~\ref{fig: hadronic e2.2/p2.2 CTA}. 

Assuming a maximum value of $f_{\rm{sc}}$= 0.01, we see that the compact jets of \gx\ can accelerate CRs up to 100\,TeV. Consequently, if this is true and moreover the entire population of BHXBs can accelerate CRs up to 100\,TeV, then BHXBs may contribute to the Galactic CR spectrum up to the knee depending on their total number \citep[see also][]{cooper2020xrbcrs}.

\subsection{Non-thermal proton power}

The uncomfortably high proton powers needed for lepto-hadronic jet models has been a topic of discussion for many years \citep[see e.g.,][]{boettcher2013leptohadronic,zdziarski2015hadronic,Liodakis2020,kantzas2020cyg}. As discussed in Section~\ref{sec: MW spectrum and dynamics}, given what we see in AGN jet observations and simulations, we would expect proton acceleration to happen primarily at the interface between the spine and the sheath of the jet, a region of limited volume \citep[][]{rieger2019particle}. In our current setup, as a first approximation, we can limit the volume where proton acceleration occurs by reducing the extent of this region with respect to the total jet length. In particular, similar to previous studies \citep[][]{Romero2008proton,Vila2010gx339,Zhang2010neutrinoLMXRBs,pepe2015lepto,Hoerbe2020importance}, we terminate the proton acceleration at a distance 10\,$z_{\rm diss}$ from the region where acceleration initiates. As a consequence, we see that even for a hard power law index of $p_{\rm p}$=1.7, the TeV emission of \gx\ due to hadronic processes will not be detectable by CTA, but the energy budget remains within reasonable values. 

A further way to constrain the total power of the accelerated protons is by increasing the minimum energy of the accelerated particles  \citep[][]{zdziarski2015hadronic,pepe2015lepto}. We nevertheless decide to use as the minimum energy for the accelerated leptons the peak of the MJ distribution and for the accelerated protons the rest mass energy (see Section~\ref{sec: model}), but will explore this in more detailed future work.

Recent high resolution magneto-hydrodynamic simulations have shown that jets can be significantly mass-loaded via instabilities at distances well beyond the launching point \citep{chatterjee2019accelerating}. This progressive mass-loading could significantly reduce the total proton power and make the hadronic models more viable, but this is a project we will pursue in the future.

\subsection{\gr\ attenuation on the optical/IR emission}
In both lepto-hadronic models, the optical emission is produced in the jet base due to synchrotron emission from the thermal leptons. The GeV-to-TeV \gr\ emission on the other hand, is produced in the particle acceleration region and above, which is located at some distance of 3000\,$r_{\rm g}$ from the black hole, two orders of magnitude further away from the jet base. Moreover, the \gr\ is beamed away making it difficult for any attenuation on this optical emission. 
The IR emission of \gx\ is produced in the particle acceleration region where the \gr\ emission originates as well. We therefore examine any \gr\ attenuation on the IR emission.

We calculate the optical depth of a 3\,TeV \gr\ that has the maximum likelihood to interact with the $\sim$0.08\,eV IR emission using equation 16 of \cite{Mastichiadis2002}: 
\begin{equation}
    \tau_{\gamma\gamma} = \frac{R_{\rm diss}}{4\pi} \int \epsilon_{\rm ph} n_{\rm ph}(\epsilon_{\rm ph}) \int d\Omega (1-\cos \theta)\, \sigma_{\gamma\gamma} \approx 10^{-8},
\end{equation}
where $\epsilon_{\rm ph}$ is the target photon energy and $n_{\rm ph}$ is the target photon number density of the particle acceleration region. 
Such a small values indicates that the particle acceleration region is optically thin to TeV \gr s.

\subsection{\gr s from BHXBs}
Despite the fact that \gx\ is considered a 'canonical' low-mass BHXB, it is also amongst the most distant ones. There are Galactic low-mass BHXBs that are as close as approximately 1--3\,kpc, e.g., 
GRO~J0422+32 \citep{webb2000gro,Gelino2003,Hynes2005},
XTE~J1118+480 \citep{Gelino2006,Hern_ndez_2008},
XTE~J1650--500 \citep{Homan2006xteJ1650,Orosz_2004},
GRO~J1655--40 \citep{hjellming1995episodic,Shahbaz1999GROJ1655,Beer2002GROJ1655},
GRS~1716--249 \citep{Remillard2006},
GS~2000+251 \citep{Casares1995,Barret_1996,Harlaftis_1996},
V404~Cyg \citep{Miller_Jones_2009},
VLA~J2130+12 \citep{Kirsten2014,Tetarenko2016OUTSIDE}, 
Swift~J1357.2--0933 \citep{Shahbaz2013,Torres2015},
MAXI~J1348--630 \citep{Chauhan2020distance}
and many more at unknown distances that might also be as low as 2--3\, kpc \citep[see][]{liu2007catalogue,Kreidberg2012,Tetarenko2016}. 

\begin{figure}
    \centering
    \includegraphics[width=1.\columnwidth]{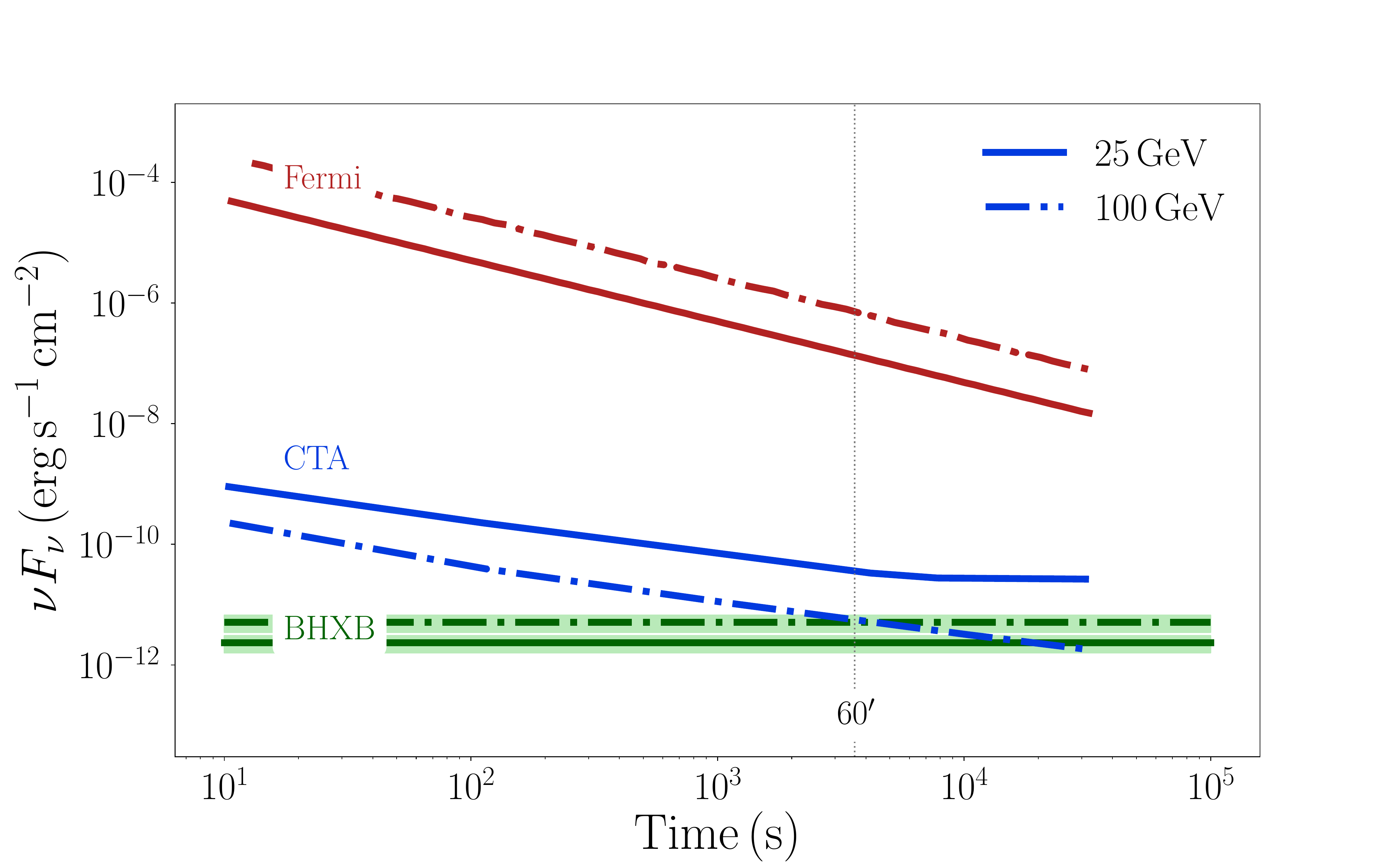}
    \caption{The \gr\ light curves in two energy bins as indicated in the legend. The horizontal green lines indicate the predicted flux of a BHXB with the same luminosity as \gx\ but located at a distance of 3\,kpc instead. We assume that the accelerated particles follow a power-law with index $p_{\rm e}$= 1.7 and $p_{\rm p}$= 1.7, and the emitted flux remains constant for one day. CTA can detect such a GeV emission within the first hour of the outburst, but \textit{Fermi}/LAT is not sensitive enough to detect such an outburst.
    }
    \label{fig: sensitivity curves hard}
\end{figure}

For this reason we also check whether some BHXBs at a distance of 3\,kpc with the same \gr\ luminosity and spectrum as \gx\ could be detected by CTA. Assuming that the jets in this putative source have identical properties to \gx , the \gr\ flux of a nearer source scales as $\left(d_{\rm GX~339-4}/d_{\rm source} \right)^2\,F_{\gamma}$, where $d_{\rm GX~339-4}$ and $d_{\rm source}$ are the distances of \gx\ and the source, respectively, and $F_{\gamma}$ is the \gr\ flux of \gx . We plot this \gr \ flux in Fig.~
\ref{fig: sensitivity curves hard} and compare it to the simulated sensitivity of CTA for various energies, as a function of observation time~\footnote{\href{http://www.cta-observatory.org/science/cta-performance/}{http://www.cta-observatory.org/science/cta-performance/}}. The energy range we study here coincides with the energy range of \textit{Fermi}/LAT which as we can see in Fig.~\ref{fig: sensitivity curves hard} is orders of magnitude less sensitive than CTA for short integration times. We assume that the \gr\ flux remains constant for up to one day and its uncertainty is of the order of 30 percent. 
We see CTA is sensitive enough to detect the 100\,GeV emission of a \gx-like source at 3\,kpc distance, with an exposure of approximately one hour, assuming the emission remains persistent for that long. Consequently, CTA should be able to detect GeV \gr s from several future bright outbursts of nearby Galactic BHXBs assuming that the accelerated particles form hard spectra within the relativistic jets produced at peak hard/hard-intermediate states.

\begin{figure}
    \centering
    \includegraphics[width=1.\columnwidth]{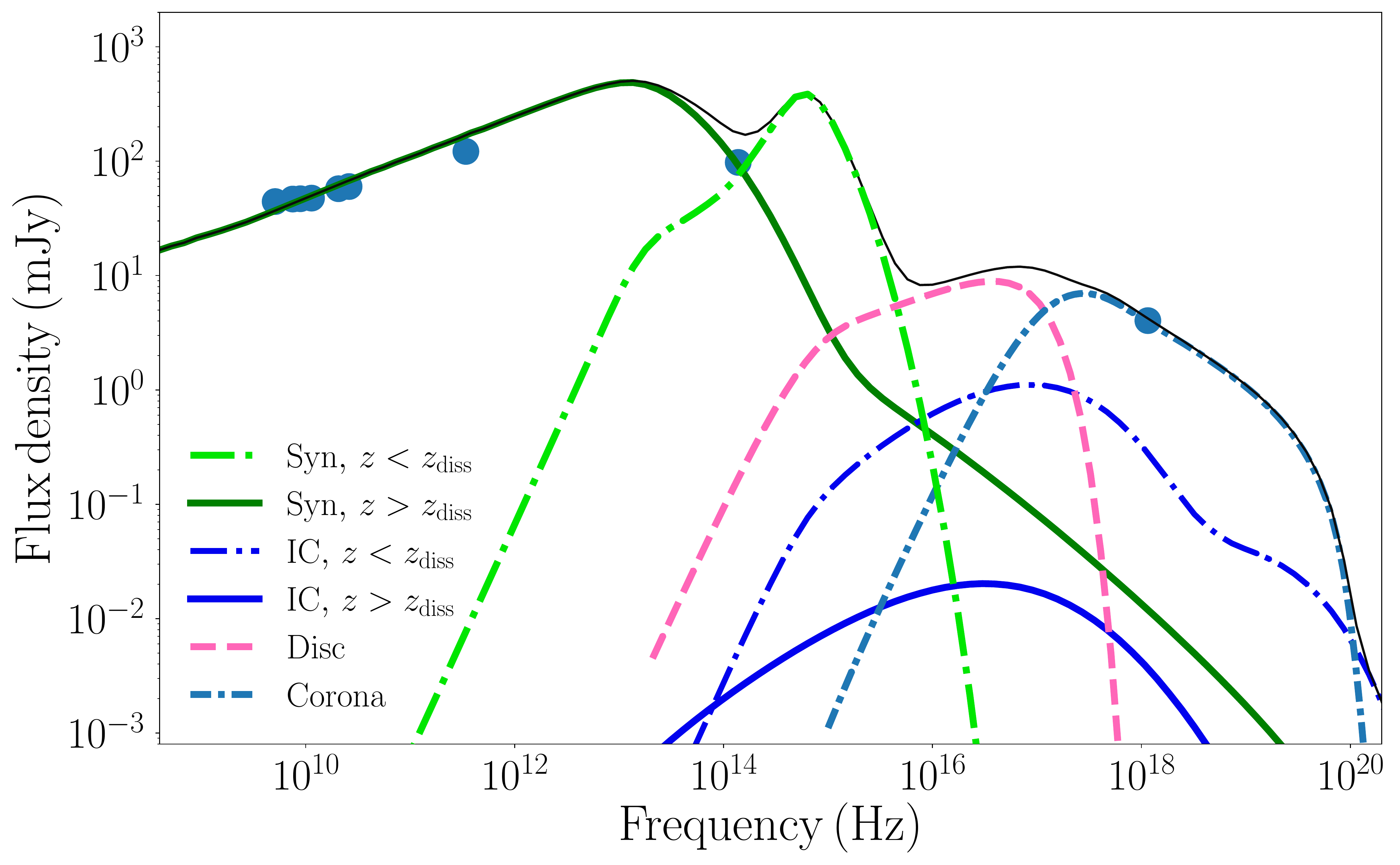}
    \caption{
        The predicted flux density based on the lepto-hadronic scenario with $p_{\rm p}$=1.7 for the 2018 outburst of MAXI~J1820+070. We compare the total emitted spectrum to the data of \protect\cite{tetarenko2021measuring}.  The rest of the components are the same as in Fig.~\ref{fig: leptonic}.}
    \label{fig: maxi j1820 mjy}
\end{figure}

\begin{figure}
    \centering
    \includegraphics[width=1.\columnwidth]{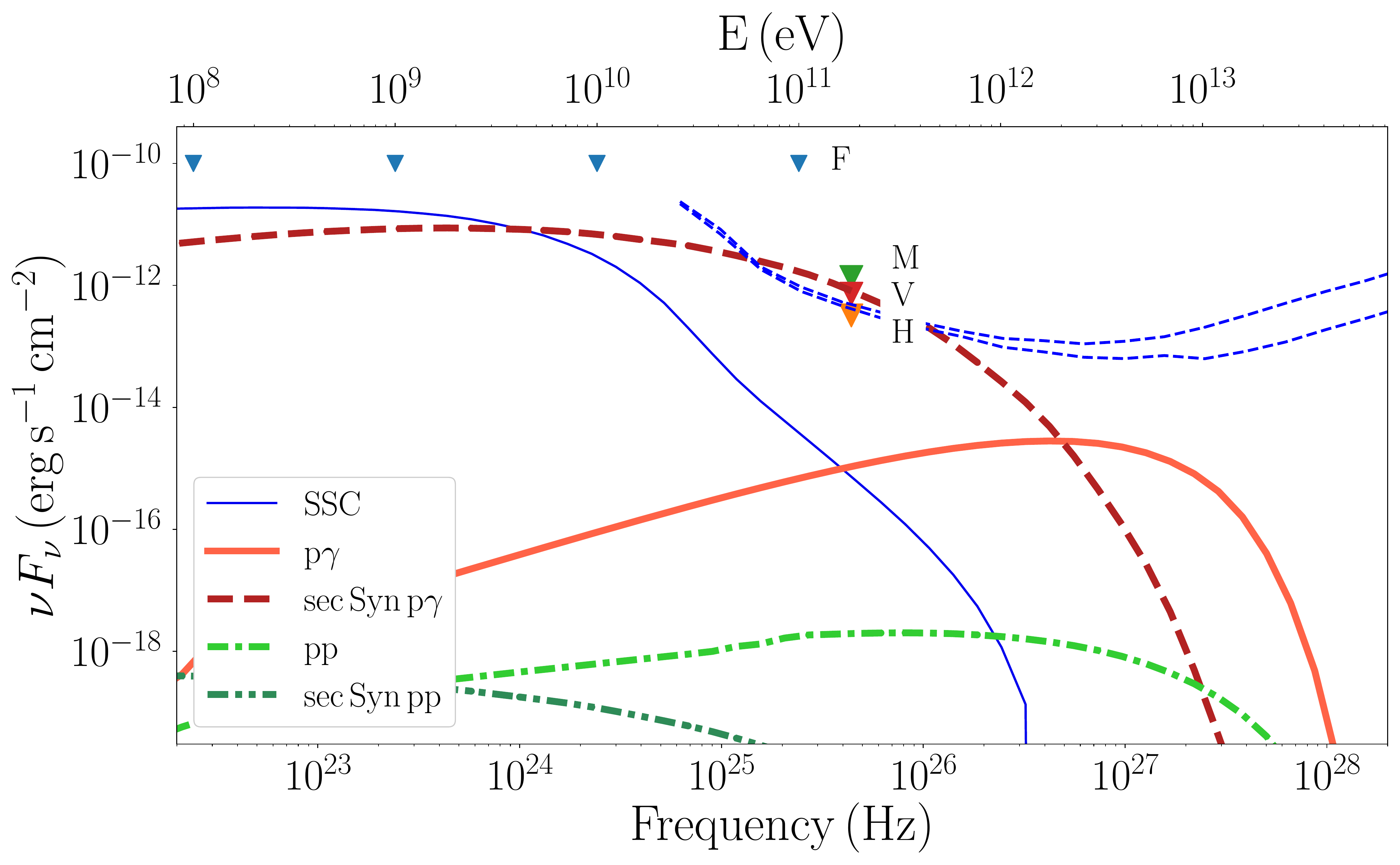}
    \caption{
        The spectral energy distribution based on the lepto-hadronic scenario with $p_{\rm p}$=1.7 for the 2018 outburst of MAXI~J1820+070. We compare the total emitted spectrum to the upper limits of \textit{Fermi}/LAT (F), MAGIC (M), VERITAS (V) and HESS (H) from \protect\cite{hoang2019multi}, and the predicted 50-hour sensitivity of the North and South site of CTA (from \href{www.cta-observatory.org}{www.cta-observatory.org}). The rest of the components are the same as in Fig.~\ref{fig: hadronic e2.2/p2.2 CTA}.}
    \label{fig: maxi j1820 sed}
\end{figure}

We finally examine a more specific example, in particular that of MAXI J1820+070, which is at 2.96\,kpc \citep{Gandhi2019Gaia,Atri2020parallax}. During its outburst in 2018, the source was monitored across the multi-wavelength spectrum, from radio to X-rays  \citep[][]{Tucker_2018}. Here, we merely benchmark the spectral energy distribution instead of optimising to determine the best fit, with the goal of illustrating the similarities and differences with our results on \gx . We use the radio-to-X-ray spectrum, as presented by \cite[][]{tetarenko2021measuring}. We set the black hole mass at 8.5\,$\rm{M_{\odot}}$ \citep{Torres_2020}, the inclination angle at 63$^{\circ}$ and the injected jet power at 15\% of the Eddington luminosity  \citep[][]{Atri2020parallax}. We take the same model parameters we found for the best fit of \gx\ for the case of $p_{\rm e} = p_{\rm p}$= 1.7 and present the spectral energy distribution of the 2018 outburst in Fig.~\ref{fig: maxi j1820 mjy} and \ref{fig: maxi j1820 sed}. We see that the radio-to-X-ray spectrum is similar to the one of \gx, namely the radio spectrum is due to non-thermal synchrotron radiation, the optical band is due to thermal synchrotron in agreement with \cite{tetarenko2021measuring} (although see \cite{Veledina2019evolving} for further contributors), and the X-ray spectrum is due to a thermal corona. In contrast to \gx, the p\g\ emission exceeds the CTA sensitivity in the sub-TeV regime. We further compare our predicted spectrum in Fig.~\ref{fig: maxi j1820 sed} to the upper limits set by Fermi/\textit{LAT} and the Cherenkov telescopes MAGIC, VERITAS and HESS \citep[][]{hoang2019multi}. We see that the predicted emission exceeds the upper limits of HESS and marginally those of VERITAS, but it is worth mentioning that these upper limits are derived after 26.9 and 12.2 hours, respectively \citep[][]{hoang2019multi}. We are unable to capture the timing signature of the TeV emission with the current version of our model, but we moreover do not know yet whether the high-energy emission of these sources is persistent for up to 20-30 hours \citep{bodaghee2013gamma}. If MAXI~J1820+070's TeV emission persists for at least a couple of hours during its next outburst, it could then be a possible target-of-opportunity for CTA. Moreover, based on the population-synthesis results of \cite{Olejak2019synthesis} and on the recent X-ray observations of \cite{Hailey2018cusp} and \cite{Mori_2021},  \cite{cooper2020xrbcrs} estimated that a few thousands BHXBs may reside in the Galactic disc capable of accelerating protons to high energy (also see \citealt{fender2005CRXRBs}). If these sources spend approximately 1\% of their outburst in the hard to hard-intermediate state \citep[][]{Tetarenko2016}, then CTA might be able to detect a few tens of BHXBs in its first years of operation.

In our current analysis, we assume equal number density of electrons and protons in the jets, similar to previous studies \citep{Vila2010gx339,Connors2019combining}. Following this assumption, we derive the jet kinetic power to be $2\times 10^{37}\rm erg\, s^{-1}$.  \cite{tetarenko2021measuring} though suggest that the jets of MAXI~J1820+070 cannot be proton dominated and constrain the ratio of protons to positrons to be $\sim 0.6$ otherwise the jet kinetic power, which they estimate to be $6\times 10^{37}\rm erg\,s^{-1}$,  may reach 18 times the accretion power. We aim to further study the impact of the pair-to-proton ratio to jet evolution and emission in a forthcoming work.

\section{Summary and conclusions}\label{sec: summary}
Astrophysical jets are ideal laboratories to understand the underlying physics of particle acceleration and the physical processes responsible for the non-thermal emission. It is still unclear whether BHXB jets can accelerate particles to high enough energy to shine in the \gr\ regime of the electromagnetic spectrum. Such emission strongly depends on the composition of the jets, which remains poorly constrained for either Galactic or extragalactic jets. A possible hadronic composition would support BHXB jets as candidate sources of Galactic CRs and shed light on this long-standing open question. Understanding the jet composition is clearly crucial not only for a better understanding of the non-thermal radiation and total power requirements, but also for our understanding of the jet launching and bulk acceleration properties. 

To further understand the properties of Galactic jets and predict any TeV signature, we studied the `canonical' low-mass BHXB \gx \ during the bright outburst of 2010. We presented the best fit of our jet model to the multiwavelength emission and found that the whole radio-to-GeV electromagnetic spectrum can be due to primary leptonic processes. To explain both the radio and the IR/Optical bands, we require a heating mechanism similar to what we see in PIC simulations \citep[][]{Sironi2009,Sironi2011,crumley2019kinetic}. We further found that the jets of \gx\ can accelerate protons to a non-thermal power law up to a few hundreds of TeV. Depending on the power-law index, we saw that the accelerated protons can produce a strong TeV emission via neutral pion decay and synchrotron radiation of secondary pairs. In the case of a hard power law of protons in particular, we found that the photomeson processes dominate the pp interactions and the synchrotron emission of secondary pairs dominates the sub-TeV band.  

\gx \ is however a distant source, located at 8\,kpc and the predicted TeV flux will not be strong enough to be detected by future \gr\ facilities, such as CTA. We rescaled the emitted spectrum to a distance of 3\,kpc and compared it to the predicted timing sensitivity of CTA. We find that CTA would be able to detect such emission with an hour of integrated observations in the energy range above 100\, GeV, which would be an indication that protons are accelerated into a hard power law. We further tested this scenario by bench-marking the electromagnetic spectrum of a nearby source, such as the newly discovered BHXB MAXI J1820+070. We found that this source might be a potential target-of-opportunity for future CTA observations to hint BHXBs as TeV sources and CR accelerators.

\section*{Acknowledgements}

We would like to thank K. Chatterjee for many fruitful conversations on jet physics and A. López-Oramas for the insightful discussion on \gr\ Astronomy. DK, SM and ML are grateful for support from the Dutch Research Council (NWO) VICI grant (no. 639.043.513). CC acknowledges support from the Swedish Research Council (VR). R.M.T.C acknowledges support from NASA grant NNG08FD60C. This research made use of \verb+ASTROPY+ (\href{http://www.astropy.org}{http://www.astropy.org}), a community-developed core \verb+PYTHON+ package for Astronomy \citep{astropy:2013, astropy:2018}, \verb+MATPLOTLIB+ \citep{Hunter:2007}, \verb+NUMPY+ \citep{oliphant2006guide}, \verb+SCIPY+ \citep{2020SciPy-NMeth}, \texttt{ISIS} functions (ISISscripts) provided by ECAP/Remeis observatory and MIT (\href{http://www.sternwarte.unierlangen.de/isis/}{http://www.sternwarte.unierlangen.de/isis/}), and the CTA instrument response functions provided by the CTA Consortium and Observatory (see \href{http://www.cta-observatory.org/science/cta-performance/}{http://www.cta-observatory.org/science/cta-performance/} for more details).

\section*{Data availability}
All observational data in this paper are publicly available (see Table~\ref{table: data}). The output of our model and the plotting scripts are available in Zenodo, at \href{https://dx.doi.org/}{https://dx.doi.org/}



\bibliographystyle{mnras}
\bibliography{GX339-4} 


\appendix

\begin{figure*}
    \centering
	\begin{minipage}{\columnwidth}
        \includegraphics[width=\columnwidth]{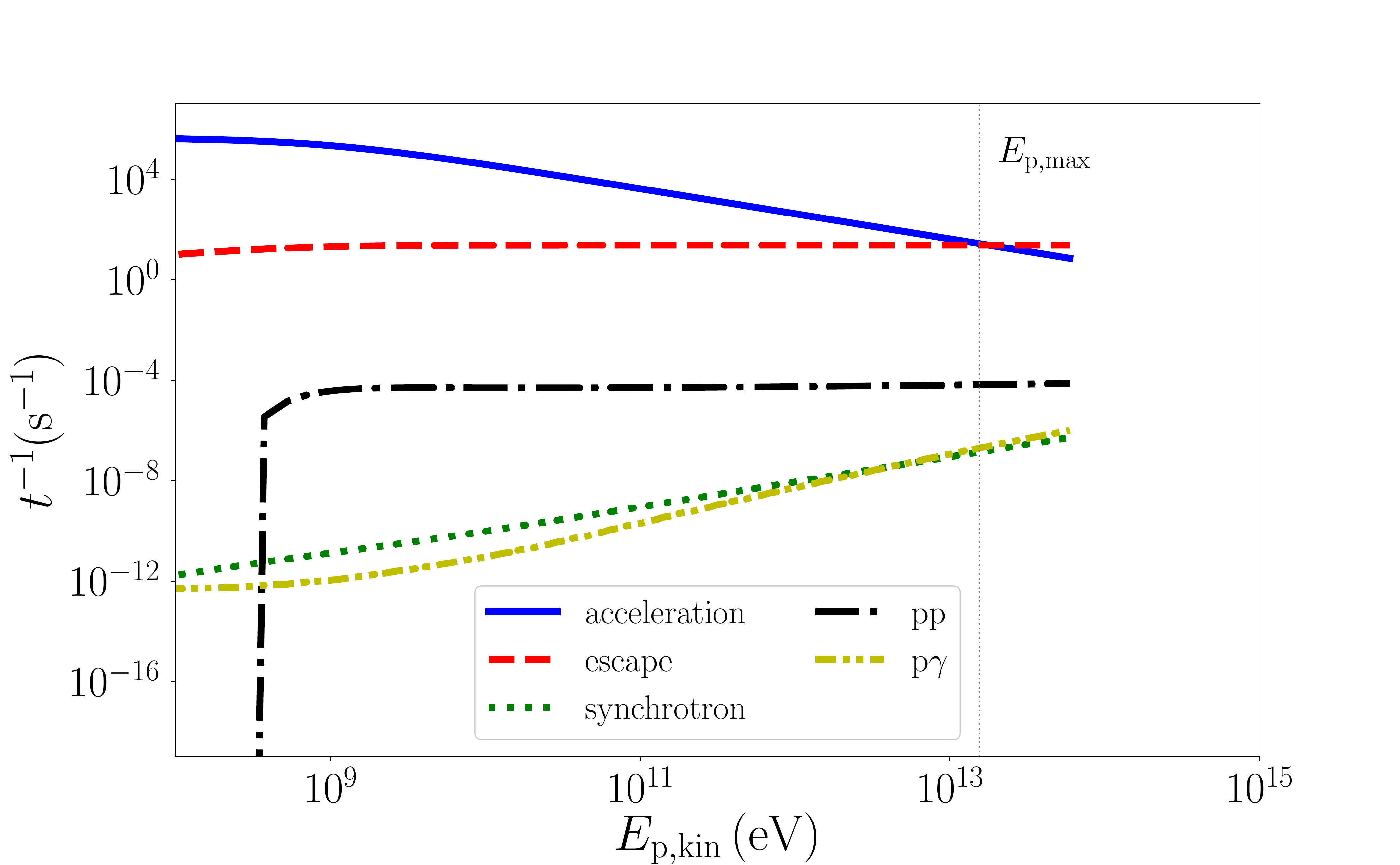}
    \end{minipage}
    \begin{minipage}{\columnwidth}
        \includegraphics[width=\columnwidth]{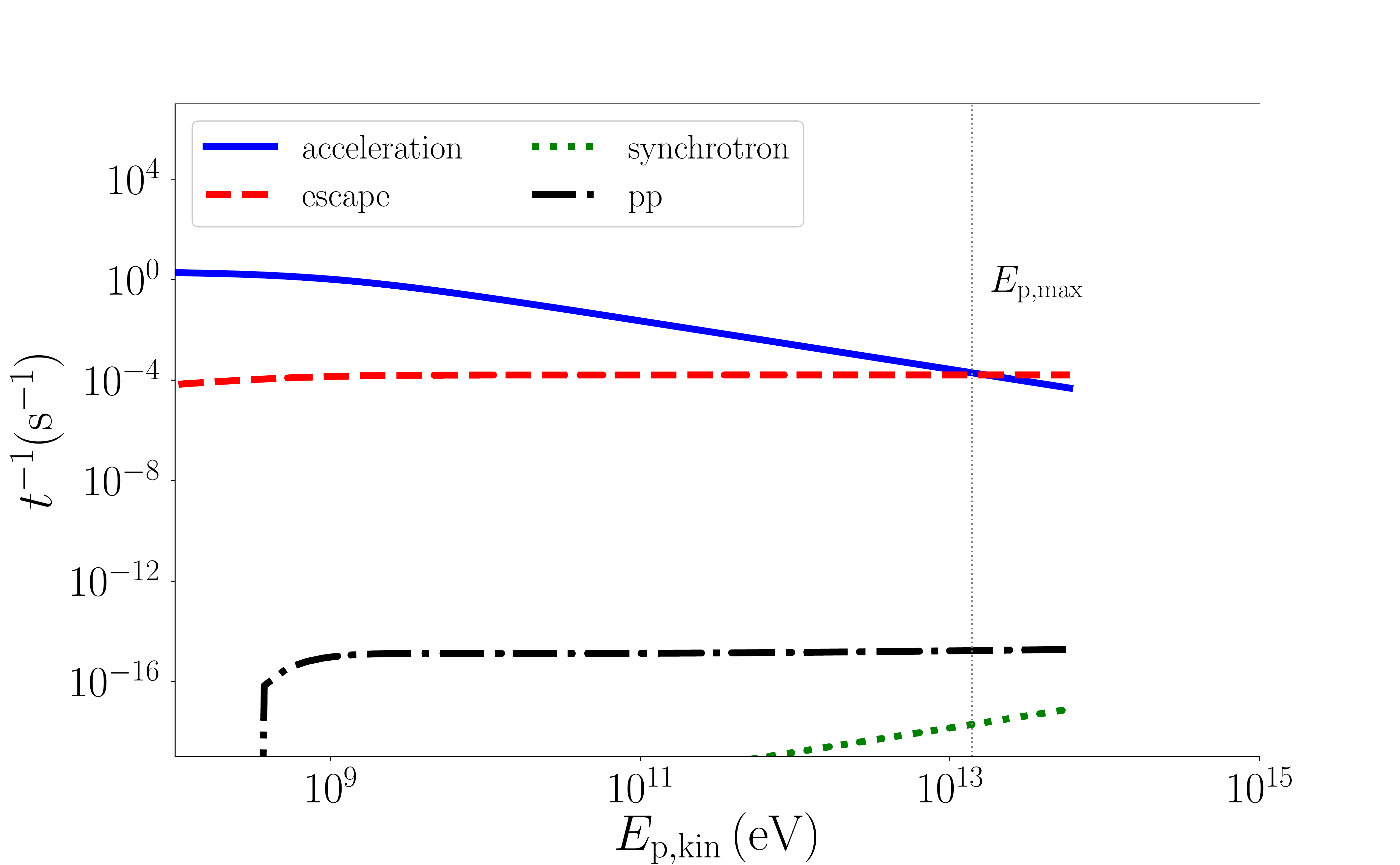}
    \end{minipage} \\
	\begin{minipage}{\columnwidth}
        \includegraphics[width=1.\columnwidth]{figs/12/hadronic/timescales/timescales_protons_zmax_had_22.pdf}
    \end{minipage}
	\begin{minipage}{\columnwidth}
        \includegraphics[width=1.\columnwidth]{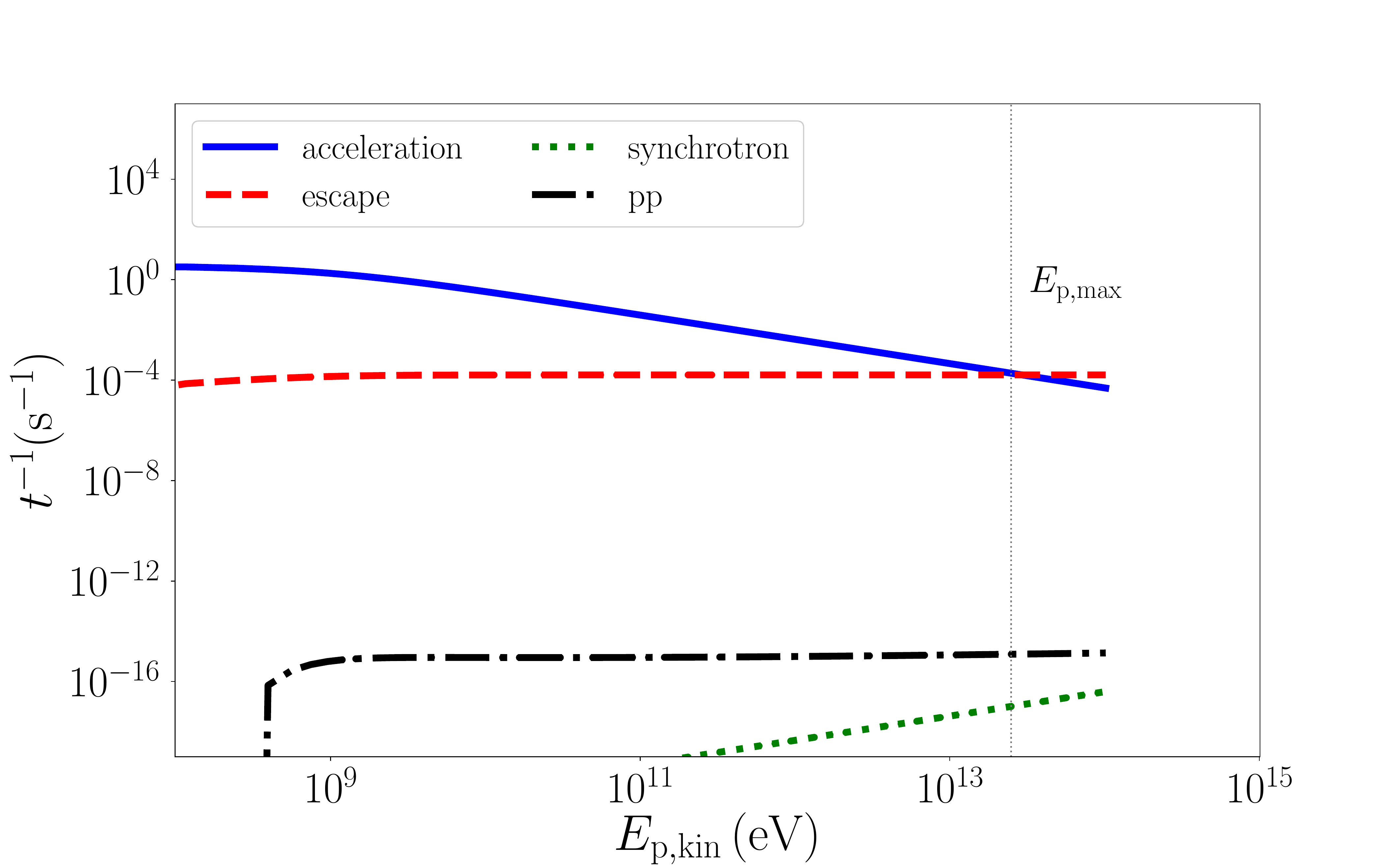}
    \end{minipage}
    \caption{The inverse of the characteristic timescales for various physical processes in the jets as indicated by the legend versus the proton kinetic energy. The top plots correspond to a power-law index of $p_{\rm p}$= 2.2 and the bottom plots correspond to a power-law index of $p_{\rm p}$= 1.7. The left plots correspond to the particle acceleration region and the right plots to the final jet segment. The vertical line in each plot shows the maximum energy  }
    \label{fig: appendix timescales}
\end{figure*}
\section{Proton characteristic timescales}\label{appendix: proton timescales}
In Fig.~\ref{fig: appendix timescales} we show the characteristic cooling timescales of synchrotron, escape, pp and p\g\ for the accelerated protons inside the jets, in comparison to the acceleration timescale. When any of the energy-loss timescales intersects the acceleration timescale we derive the maximum proton energy for every jet segment. The residence timescale is the one that defines the maximum proton energy inside the jets.

\bsp	
\label{lastpage}
\end{document}